\documentclass[twocolumn,prl,amsmath,amssymb,floatfix,superscriptaddress,showpacs,longbibliography]{revtex4-1}
\usepackage{graphicx}
\usepackage{dcolumn}
\usepackage{bm}
\usepackage{array}
\usepackage{booktabs}
\usepackage[flushleft]{threeparttable}
\usepackage{xspace}
\usepackage{hyperref}
\usepackage{color}
\def\new{\color{black}}

\def\rucl{$\alpha$-RuCl$_3$\xspace}
\UseRawInputEncoding

\begin{document}

\title{Neutron spectroscopy evidence for {\new a possible} magnetic-field-induced gapless quantum-spin-liquid phase in a Kitaev material $\alpha$-RuCl$_3$}

\author{Xiaoxue~Zhao（赵晓雪）}
\altaffiliation{These authors contributed equally to the work.}
\affiliation{National Laboratory of Solid State Microstructures and Department of Physics, Nanjing University, Nanjing 210093, China}
\author{Kejing~Ran（冉柯静）}
\altaffiliation{These authors contributed equally to the work.}
\author{Jinghui~Wang（王靖珲）}
\affiliation{School of Physical Science and Technology and ShanghaiTech Laboratory for Topological Physics, ShanghaiTech University, Shanghai 200031, China}
\author{Song~Bao（鲍嵩）}
\author{Yanyan~Shangguan（上官艳艳）}
\affiliation{National Laboratory of Solid State Microstructures and Department of Physics, Nanjing University, Nanjing 210093, China}
\author{Zhentao~Huang（黄振涛）}
\author{Junbo~Liao（廖俊波）}
\author{Bo~Zhang（张波）}
\author{Shufan~Cheng（承舒凡）}
\author{Hao~Xu（徐豪）}
\affiliation{National Laboratory of Solid State Microstructures and Department of Physics, Nanjing University, Nanjing 210093, China}
\author{Wei~Wang（王巍）}
\affiliation{School of Science, Nanjing University of Posts and Telecommunications, Nanjing 210023, China}
\author{Zhao-Yang~Dong（董召阳）}
\affiliation{Department of Applied Physics, Nanjing University of Science and Technology, Nanjing 210094, China}
\author{Siqin~Meng（孟思勤）}
\affiliation{Helmholtz-Zentrum Berlin f\"{u}r Materialien und Energie GmbH, Hahn-Meitner-Platz 1D-14109 Berlin, Germany}
\affiliation{China Institute of Atomic Energy, Beijing 102413, China}
\author{Zhilun~Lu（陆智伦）}
\affiliation{Helmholtz-Zentrum Berlin f\"{u}r Materialien und Energie GmbH, Hahn-Meitner-Platz 1D-14109 Berlin, Germany}
\affiliation{School of Engineering and the Built Environment, Edinburgh Napier University, Edinburgh EH10 5DT, United Kingdom}
\author{Shin-ichiro~Yano}
\affiliation{National Synchrotron Radiation Research Center, Hsinchu 30077, Taiwan}
\author{Shun-Li~Yu（于顺利）}
\email{slyu@nju.edu.cn}
\author{Jian-Xin~Li（李建新）}
\email{jxli@nju.edu.cn}
\author{Jinsheng~Wen（温锦生）}
\email{jwen@nju.edu.cn}
\affiliation{National Laboratory of Solid State Microstructures and Department of Physics, Nanjing University, Nanjing 210093, China}
\affiliation{Collaborative Innovation Center of Advanced Microstructures, Nanjing University, Nanjing 210093, China}

\begin{abstract}

As one of the most promising Kitaev quantum-spin-liquid (QSL) candidates, $\alpha$-RuCl$_3$ has received a great amount of attention. However, its ground state exhibits a long-range zigzag magnetic order, which defies the QSL phase. Nevertheless, the magnetic order is fragile and can be completely suppressed by applying an external magnetic field. Here, we explore the evolution of magnetic excitations of $\alpha$-RuCl$_3$ under an in-plane magnetic field, by carrying out inelastic neutron scattering measurements on high-quality single crystals. Under zero field, there exist spin-wave excitations near the $M$ point and a continuum near the $\mit\Gamma$ point, which are believed to be associated with the zigzag magnetic order and fractional excitations of the Kitaev QSL state, respectively. By increasing the magnetic field, the spin-wave excitations gradually give way to the continuous excitations. On the verge of the critical field $\mu_0H_{\rm c}=7.5$~T, the former vanish and only the latter is left, indicating the emergence of a pure QSL state. By further increasing the field strength, the excitations near the $\mit\Gamma$ point become more intense. By following the gap evolution of the excitations near the $\mit\Gamma$ point, we are able to establish a phase diagram composed of three interesting phases, including a gapped zigzag order phase at low fields, {\new possibly-}gapless QSL phase near $\mu_0H_{\rm c}$, and gapped partially polarized phase at high fields. These results demonstrate that an in-plane magnetic field can drive $\alpha$-RuCl$_3$ into a long-sought QSL state near the critical field.
\end{abstract}


\maketitle
In the past few years, \rucl with the honeycomb structure has been studied extensively in the pursuit of Kitaev quantum spin liquids (QSLs), which are resulting from the bond-dependent anisotropic Kitaev interactions\cite{aop321_2,0034-4885-80-1-016502,nrp1_264,npjqm4_12}, different from the triangular-, kagome-, or pyrochlore-structured QSL candidates with geometrical frustration\cite{Anderson1973153,nature464_199}. Now, it is well established that the ground state of \rucl is actually a zigzag ordered state\cite{PhysRevB.91.144420,PhysRevB.91.241110,PhysRevB.92.235119,nm15_733,Ritter_2016}. However, taking advantage of the spatial anisotropy of the Ru$^{3+}$ $d$ orbitals and the close-to-ideal bond configurations, it has been shown that there exists a large Kitaev interaction between the effective spin-1/2 moments\cite{PhysRevB.90.041112,PhysRevB.93.155143,sr6_37925,PhysRevLett.118.107203,PhysRevB.96.115103,np16_837}. Due to the presence of the Kitaev interaction, the zigzag order phase is in proximity to the Kitaev QSL phase\cite{PhysRevLett.114.147201,nm15_733,np12_912,np13_1079}, although there also exist some non-Kitaev terms that make the system deviate from the QSL phase\cite{PhysRevB.96.054410,PhysRevB.98.100403,PhysRevB.100.075110,prl105_027204,PhysRevLett.112.077204,PhysRevB.92.024413,1367-2630-16-1-013056,PhysRevLett.113.107201,np11_462,prl102_017205,0953-8984-29-49-493002,PhysRevB.93.214431}. This provides the opportunity that by tuning the competing interactions, the zigzag magnetic order can be suppressed and a QSL state may be achieved\cite{PhysRevB.99.140413,PhysRevB.96.064430,PhysRevB.102.140402,nc10_2470,PhysRevLett.120.077203,Janssen_2019,PhysRevResearch.2.013014,nc12_4512,nc12_4007}. In fact, there are accumulating reports that an external magnetic field applied within the honeycomb plane can suppress the magnetic order effectively and drive the system into a magnetically disordered state, utilizing various experimental probes, including magnetization\cite{PhysRevB.91.094422,PhysRevLett.119.227208,PhysRevB.92.235119,PhysRevB.91.180401,PhysRevB.103.174417}, specific heat\cite{PhysRevB.91.094422,PhysRevLett.119.227208,PhysRevB.95.180411,PhysRevLett.125.097203,PhysRevB.96.041405,PhysRevLett.119.037201,PhysRevLett.120.067202,PhysRevB.103.054440,PhysRevB.99.094415,tanaka2022thermodynamic,zhou2022intermediate}, neutron scattering\cite{PhysRevB.92.235119,PhysRevB.95.180411,npjqm3_8,PhysRevB.100.060405,PhysRevB.103.174417}, nuclear magnetic resonance\cite{PhysRevLett.119.037201,PhysRevLett.119.227208,PhysRevB.101.020414,np14_786}, thermal conductivity and thermal Hall conductivity\cite{PhysRevLett.118.187203,PhysRevLett.120.117204,PhysRevLett.120.067202,nature559_227,PhysRevLett.120.217205,np17_915,doi:10.1126/science.aay5551,bruin2021robustness,czajka2022planar}, Raman, microwave, and terahertz spectroscopy\cite{PhysRevB.101.140410,PhysRevB.98.184408,PhysRevLett.119.227201,PhysRevLett.119.227202,PhysRevB.98.184408,PhysRevB.98.094425,wulferding2020magnon}, magnetodielectric\cite{PhysRevB.95.245104}, magnetic torque\cite{PhysRevLett.118.187203}, resonant torsion magnetometry\cite{modic2021scale}, electron spin resonance\cite{PhysRevB.96.241107,PhysRevLett.125.037202}, and thermal expansion and magnetostriction measurements\cite{PhysRevB.101.245158,PhysRevB.102.214432}. Nevertheless, whether the disordered phase under field is the long-sought QSL phase\cite{PhysRevLett.120.067202,PhysRevB.103.054440,PhysRevB.99.094415,tanaka2022thermodynamic,PhysRevLett.120.117204,nature559_227,PhysRevLett.120.217205,np17_915,doi:10.1126/science.aay5551,bruin2021robustness,czajka2022planar,PhysRevB.101.245158,np14_786,npjqm3_8,PhysRevB.100.060405,PhysRevB.101.020414,PhysRevLett.118.187203,PhysRevB.98.184408,PhysRevB.101.140410,wulferding2020magnon,PhysRevResearch.2.033011,modic2021scale,PhysRevB.102.214432,PhysRevB.96.241107,PhysRevLett.125.037202}, and if it is, whether it is gapless or gapped\cite{PhysRevLett.119.227208,PhysRevB.95.180411,PhysRevLett.119.037201,npjqm3_8,PhysRevB.101.020414}, remain hotly debated. Furthermore, there are also some controversies on whether the field divides the phase diagram into two parts, or three parts with a QSL phase intermediate between the low- and high-field phases\cite{PhysRevB.96.041405,PhysRevLett.119.037201,PhysRevLett.120.067202,PhysRevB.103.054440,tanaka2022thermodynamic,npjqm3_8,PhysRevB.100.060405,PhysRevB.101.020414,np14_786,PhysRevLett.118.187203,PhysRevLett.120.117204,PhysRevB.95.180411,PhysRevLett.125.097203,nature559_227,PhysRevLett.120.217205,np17_915,doi:10.1126/science.aay5551,bruin2021robustness,PhysRevLett.119.227208,modic2021scale,PhysRevB.96.241107,PhysRevLett.125.037202,PhysRevB.101.245158,PhysRevB.102.214432,PhysRevResearch.2.033011}.    

In this Letter, we aim to solve these problems by carrying out inelastic neutron scattering (INS) measurements on the magnetic field evolution of the magnetic excitations {\new with finer field step of 0.5~T, higher energy resolution of 0.15~meV, and stronger field strength up to 13~T, as compared to previous INS works under fields\cite{npjqm3_8,PhysRevB.100.060405}.} Under zero field, the magnetic excitations are composed of the spin-wave excitations associated with the zigzag magnetic order\cite{nm15_733,PhysRevLett.118.107203,Banerjee1055}, and a continuum hypothesized to be the fractional excitations associated with the Kitaev QSL state\cite{PhysRevLett.114.147201,Kejing_Ran:27501,np12_912,np13_1079,nm15_733}, which are around the $M$ and $\mit\Gamma$ points, respectively. Under an external magnetic field applied within the $a$-$b$ plane, the spin-wave excitations around the $M$ point are gradually suppressed and vanish around the critical field $\mu_0H_{\rm c}\approx7.5$~T, accompanying the suppression and disappearance of the zigzag magnetic order. On the other hand, the continuum near the $\mit\Gamma$ point still persists when the spin waves vanish. These results are evident that the continuum around the $\mit\Gamma$ point represents the fractional excitations associated with the QSL state, and the phase near $\mu_0H_{\rm c}$ is the QSL phase. By following the gap evolution of the continuum, we can divide the phase diagram into three phases, including the low-field gapped zigzag ordered state, intermediate {\new possibly-}gapless QSL, and gapped partially polarized phase.

\begin{figure}[ht]
\centering
\includegraphics[width=0.85\linewidth]{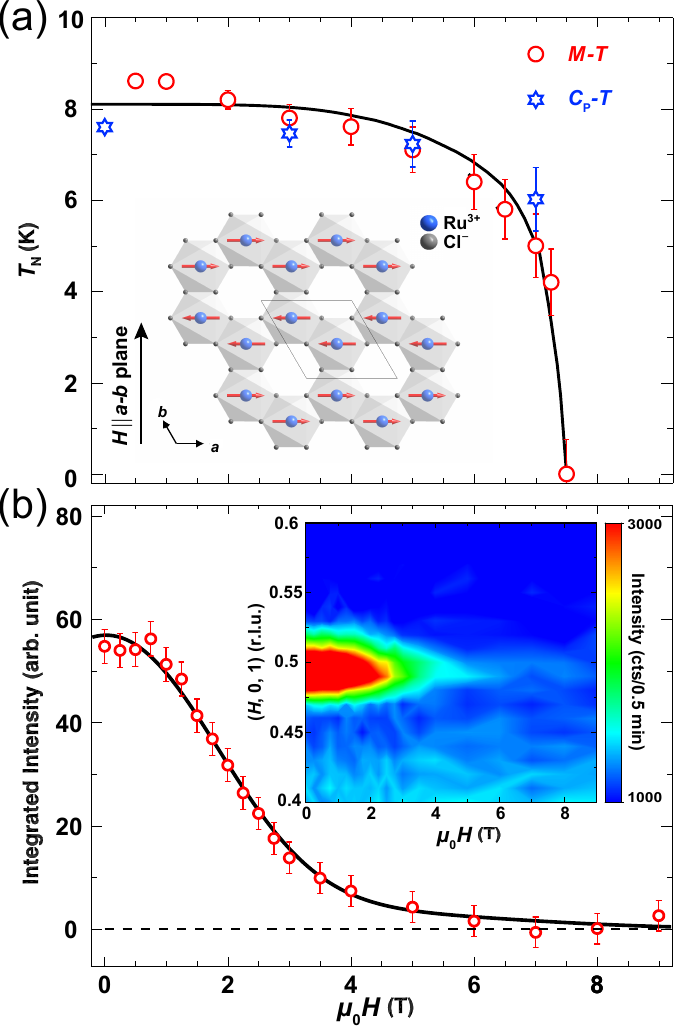}
\caption{\label{fig1}{(a) Dependence of the magnetic transition temperature $T_{\rm N}$ as a function of the in-plane magnetic field $\mu_0H$ for \rucl, obtained from the magnetization and specific heat data. The inset shows the schematic honeycomb crystal lattice of \rucl with the zigzag magnetic order. (b) Field dependence of the integrated intensities of the magnetic Bragg peak (0.5,\,0,\,1). The inset is a contour map showing the elastic scans through (0.5,\,0,\,1), under magnetic field applied along the [-1,\,2,\,0] direction with strength ranging from 0 to 9~T. Black solid curves through data are guides to the eye. The errors represent one standard deviation throughout the paper.}} 
\end{figure}

\begin{figure*}
\centering
\includegraphics[width=0.95\linewidth]{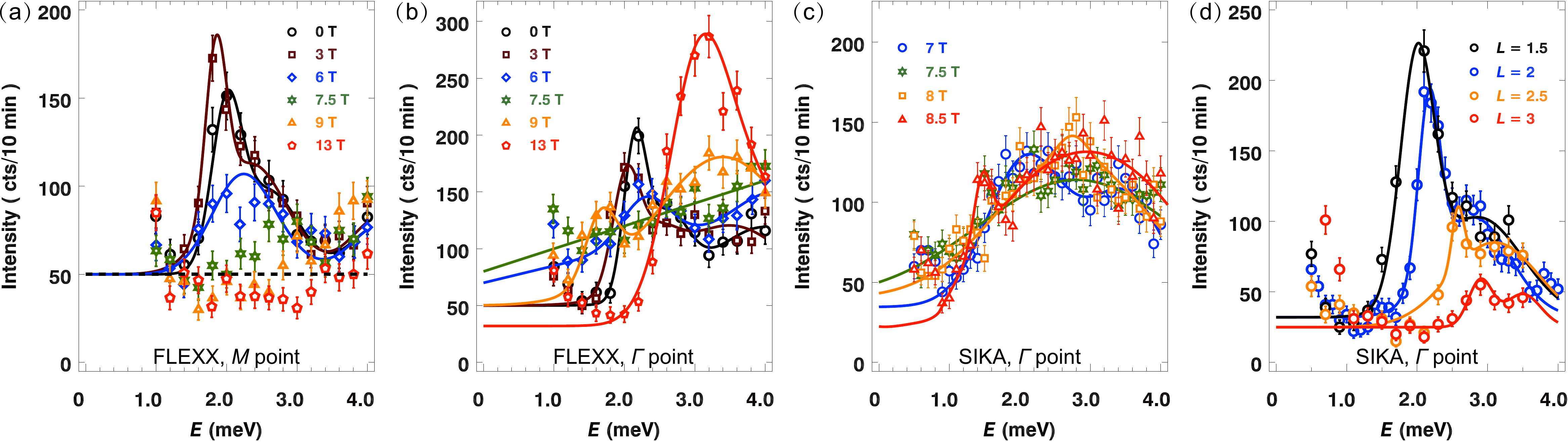}\\
\caption{\label{fig2}{(a) Constant-$\bm{Q}$ scans at the $M$ point (0.5,\,0,\,2) of the two-dimensional Brillouin zone, with applied field up to 13~T. (b) Same as (a) but at the Brillouin zone center $\mit\Gamma$ point (0,\,0,\,2). {\new (c) Similar scans as those in (b) but with finer field step of 0.5~T ranging from 7 to 8.5~T, around the critical field of 7.5~T. (d)} Constant-$\bm{Q}$ scans at the $\mit\Gamma$ point (0,\,0,\,$L$) with different $L$s under zero field. All measurements were performed on Sample I at $T=1.8$~K. (a) and (b) were both measured on FLEXX triple-axis spectrometer while (c) {\new and (d) were} measured on SIKA. Solid lines are guides to the eye, and black dotted horizontal lines represent background signals.}}
\end{figure*}

Single crystals of \rucl were grown by the chemical vapor transport method using commercially-purchased anhydrous \rucl powders\cite{PhysRevLett.118.107203,Kejing_Ran:27501}. The plate-like crystals are shiny and black with a typical size of 60~mg for each piece. Magnetic susceptibility measurements were performed using the vibrating sample magnetometer option integrated in a Physical Property Measurement System (PPMS-9T) from Quantum Design. The results showed that the sample had a single magnetic transition temperature. The specific heat measurements were also conducted on PPMS-9T. Neutron scattering measurements were conducted on two cold-neutron triple-axis spectrometers, FLEXX located at Helmholtz-Zentrum Berlin (HZB), and SIKA located at Australian Nuclear Science and Technology Organization (ANSTO)\cite{Wu_2016}, both utilizing a fixed-final-energy mode with $E_{\rm f}=5.0$~meV under double-focusing conditions for both the monochromator and analyzer. {\new The energy resolutions for both instruments were $\sim$0.15~meV~(full width at half maximum).} Two batches of samples, both weighed $\sim$2~g in total, were labeled as Sample I and Sample II. Sample I and II arrays consisted of 20 and 22 pieces of single crystal, respectively. The former was used for measurements on both FLEXX and SIKA, whilst the latter was only measured on SIKA. They were coaligned using a backscattering Laue x-ray diffractometer and glued onto aluminum plates by hydrogen-free Cytop grease. These crystals were well aligned so that the overall mosaic spreads were both less than 3$^\circ$, as determined from the rocking scans through the (0,\,0,\,3) and (1,\,0,\,0) Bragg peaks. All measurements were carried out in the ($H,\,0,\,L$) plane with magnetic field applied along the [-1,\,2,\,0] direction. A hexagonal structure, with the routinely-adopted lattice parameters $a=b=5.96$~\AA, and $c=17.20$~\AA, was used throughout this Letter. The wave vector $\bm{Q}$ was expressed as ($H,\,K,\,L$) in reciprocal lattice unit (r.l.u.) of $(a^{*}, b^{*}, c^{*}) = (4\pi/\sqrt3a, 4\pi/\sqrt3b, 2\pi/c)$.

\begin{figure*}[htb]
\centering
\includegraphics[width=1\linewidth]{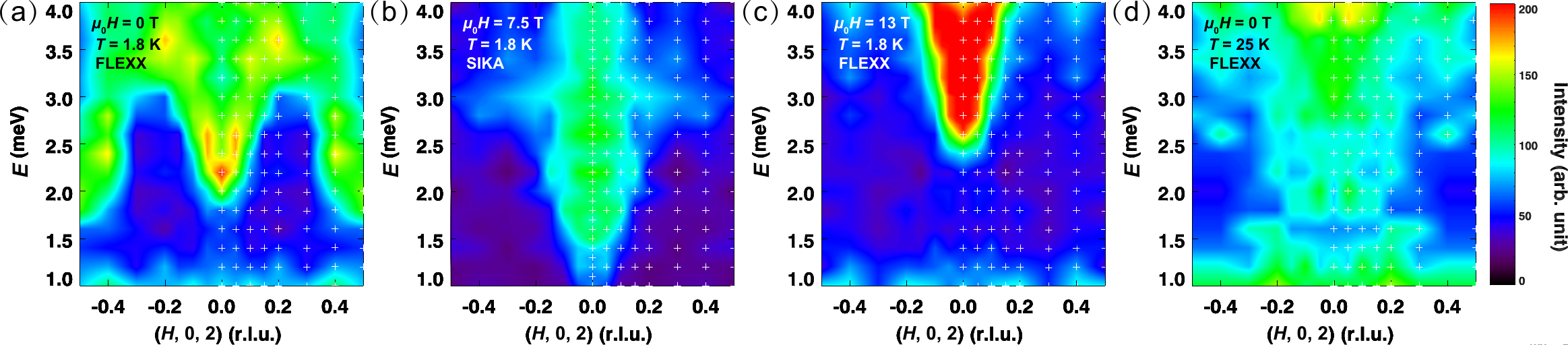}
\caption{\label{fig3}{Magnetic excitation spectra along the [100] direction under different magnetic fields. (a)-(c) were measured at 1.8~K, while (d) was measured at 25~K. Data of (b) were obtained from SIKA, and the rest were from FLEXX. $H=0$ and 0.5~r.l.u. correspond to the high-symmetric $\mit\Gamma$ and $M$ points, respectively. White cross dots mark experimental data points. In all the panels, left regions (-0.5\textless $H$\textless 0~r.l.u.) are symmetrized from the right (0\textless $H$\textless 0.5 r.l.u.) for better visualizing purpose. {\new Note that according to the linear scans in Fig.~\ref{fig2}, there are still finite intensities below 1.5~meV at the $\mit\Gamma$ point at 7.5~T, so the data in panel (b) should not be misinterpreted as there is a gap below 1.5~meV.}}} 
\end{figure*}



From susceptibility and specific heat measurements on \rucl, the relation between the magnetic transition temperature $T_{\rm N}$ and the applied in-plane magnetic field $\mu_0H$ is obtained, as presented in Fig.~\ref{fig1}(a). It clearly shows that $T_{\rm N}$ is decreasing with increasing field, indicating that the magnetic order is weakened and disappears at $\mu_0H_{\rm c}=7.5$~T. The magnetic field dependence of the integrated intensities for the magnetic Bragg peak (0.5,\,0,\,1) by elastic neutron scattering measurements is plotted in Fig.~\ref{fig1}(b). With the gradual increase of external field strength, the intensities of the Bragg peak are reduced correspondingly, also implying that the magnetic ordering is being suppressed and ultimately vanishes at around 7~T. These results are consistent with previous reports that an in-plane external field will suppress the magnetic order of $\alpha$-RuCl$_3$\cite{PhysRevB.91.094422,PhysRevLett.119.227208,PhysRevB.103.174417,PhysRevB.92.235119,PhysRevB.91.180401}.

Based on some previous experiments\cite{nm15_733,PhysRevLett.118.107203,Banerjee1055,np13_1079,Kejing_Ran:27501}, the magnetic excitations are basically converged at the $M$ and $\mit\Gamma$ points in the two-dimensional Brillouin zone. Specifically, the gapped sharp excitations around the $M$ point are the spin-wave excitations ascribed to the zigzag magnetic order, while the ones around the $\mit\Gamma$ point exhibiting broad continuous characteristics are suggested to be the fractional magnetic excitations bestowed by the proximity to the Kitaev QSL state of $\alpha$-RuCl$_3$\cite{PhysRevLett.118.107203,PhysRevLett.114.147201,Kejing_Ran:27501,np13_1079,nm15_733}. To observe the evolutions of these two types of excitations with respect to the applied field, we thus performed constant-$\bm{Q}$ scans at the $M$ point (0.5,\,0,\,2) and $\mit\Gamma$ point (0,\,0,\,2). Some of the representative data are shown in Fig.~\ref{fig2}. In Fig.~\ref{fig2}(a), it shows that the excitations at the $M$ point are enhanced from 0 to 3~T, which may be due to the spectral weight transfer of the magnetic Bragg peak---as shown in Fig.~\ref{fig1}(b), the peak intensity has a great drop from 0 to 3~T. With the field further increasing, the intensity is suppressed and almost vanishes at 7.5~T and above. As shown in Fig.~\ref{fig2}(b), the excitations at the $\mit\Gamma$ point are gradually suppressed with the field for $\mu_0H\le7.5$~T, similar to those at the $M$ point. More importantly, {\new at 7.5~T,} the peak feature of the excitations at the $\mit\Gamma$ point disappears, resulting in a featureless continuous profile in energy expected for a QSL. At 9 and 13~T, the intensities become stronger, while there are no magnetic scattering intensities at the $M$ point for the field strength exceeding 7.5~T, indicating that the system enters a partially-polarized state{\new, in which the magnetic moments are forced to be partially aligned with field.} {\new Furthermore, by more detailed measurements around the critical field of 7.5~T, as shown in Fig.~\ref{fig2}(c), we can find the peak feature of the excitations at the $\mit\Gamma$ point reappears after 8~T. Overall, the featureless and continuous scan profiles at 7.5 and 8~T are quite distinct from others. At other fields, the scan reaches its background $\gtrsim$1~meV, below which the intensity raises due to the incoherent elastic scattering. On the other hand, the data points for 7.5 and 8~T are well above the background level at 1~meV, and remain finite when extended to zero energy. From these results, we can judge that the excitations are gapless at 7.5 and 8~T, but gapped at other fields, at least on a qualitative level. This issue will be discussed further in the latter part of this work.}

In most previous neutron scattering experiments on \rucl, the $L$-dependence of the continuous excitations at the $\mit\Gamma$ point is typically not taken into account, and the models are normally based on the two-dimensional magnetic structure, ignoring the interlayer interactions. 
In Ref.~\onlinecite{PhysRevB.100.060405}, it has been pointed out that the excitations at the $\mit\Gamma$ point actually follow a cosine form, indicating a nonnegligible interlayer magnetic coupling. We have also measured the excitations at $M$ and $\mit\Gamma$ points with different $L$s. In Fig.~\ref{fig2}(d), we show four representative scans at different $L$s at zero field. The data show apparent $L$ dependence, consistent with Refs.~\onlinecite{PhysRevB.100.060405,PhysRevB.103.174417}. Nevertheless, the evolution of the excitations with the field is similar for different $L$s, as shown in the Supplementary Materials\cite{sm}.

To better visualize the evolutions of the magnetic excitations with the field, we have performed a series of energy scans at various $\bm{Q}$ values, and obtained the excitation spectra along the [100] direction under different field strengths at the base temperature as plotted in Fig.~\ref{fig3}. For comparison, the spectra at 25~K, well above the $T_{\rm N}$, are also plotted. The data in Fig.~\ref{fig3}(b) are from SIKA, and the rest are all from FLEXX. Since the measurements were carried out on similar instruments with the same Sample I, the results are similar and comparable (See Supplementary Materials for more details\cite{sm}). Figure~\ref{fig3}(a) clearly shows that there are two types of excitations---the spin wave excitations, with an energy gap around 1.6~meV, disperse upwards from the $M$ point ($H=\pm0.5$~r.l.u.), and reach the band top at the $\mit\Gamma$ point; at the $\mit\Gamma$ point, there is another type of excitations. Although they seem to have a dispersion similar to the spin waves, they are shown to be incompatible with the spin waves but are a continuum representing the fractional excitations resulting from the QSL phase instead\cite{Kejing_Ran:27501}. Intriguingly, as shown in Fig.~\ref{fig3}(b), at the critical field of $\mu_0H_{\rm c}=7.5$~T, while the spin-wave excitations at the $M$ point completely disappear, the continuum at the $\mit\Gamma$ point persists. This strongly indicates that the disordered phase at $\mu_0H_{\rm c}$ is the long-sought QSL phase featuring fractional excitations. Furthermore, the excitations appear to extend below 1~meV and become gapless. Compared Fig.~\ref{fig3}(b) with the spectra in the paramagnetic phase shown in Fig.~\ref{fig3}(d), while it is similar that the spin waves both disappear, and both feature a {\new possibly} gapless continuum, the two spectra show clear differences in their boundaries of the continua, suggesting that the field-induced {\new possibly-}gapless QSL is distinct from the paramagnetic phase above $T_{\rm N}$. By further increasing the field up to 13~T, the spin waves at the $M$ point are still absent. On the other hand, the gap at the $\mit\Gamma$ point reopens and becomes larger than that at zero field. Furthermore, the intensities of the excitations also become more intense. {\new The gapped excitations dispersing from the $\mit\Gamma$ point look like gapped ferromagnetic spin excitations. However, the excitations of a magnetic-field-driven partially polarized state may not be necessarily consistent with those of a ferromagnetic ground state under zero field, as the excitations are dependent on the Hamiltonians. For example, in our recent work on YbZnGaO$_4$, we have found that in the fully polarized state the excitations disperse from the $M$ point as expected for antiferromagnetic excitations, instead of from the $\mit\Gamma$ point as expected for ferromagnetic excitations, because of the presence of dominant antiferromagnetic interactions in YbZnGaO$_4$ under zero field\cite{PhysRevB.104.224433}. Returning to the case of \rucl, we think that similar excitations of the partially-polarized and ferromagnetic states indicate the presence of a dominant ferromagnetic Kitaev interaction, as suggested in previous works\cite{np16_837,PhysRevLett.118.187203,PhysRevB.96.115103,Kejing_Ran:27501}.}

\begin{figure}[htb]
  \centering
  \includegraphics[width=0.95\linewidth]{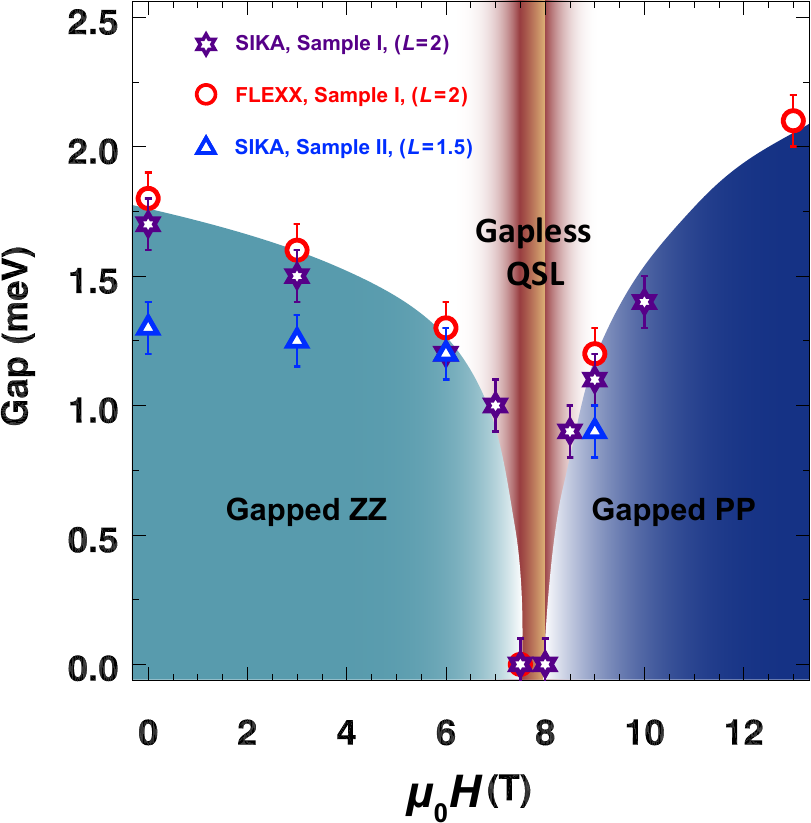}\\
  \caption{\label{fig4}{Field-evolution of the energy gap and the phase diagram. The left cyan zone denotes the gapped zigzag ordered state (Gapped ZZ), {\new central brown column represents the {\new possibly} gapless QSL state, and the right dark blue zone} is the gapped partially-polarized state (Gapped PP). The purple stars and red circles denote the data measured on Sample I with $L=2$~r.l.u. obtained from SIKA and FLEXX, respectively. The dark blue triangles represent the data measured on SIKA on Sample II with $L=1.5$~r.l.u..}}
\end{figure}

Summarizing the results in Figs.~\ref{fig2} and \ref{fig3}, and Supplementary Figs.~1 and 2, we can obtain a phase diagram of \rucl based on the gap size in Fig.~\ref{fig4}. Here, we define the energy gap as the energy where the intensity starts to rise in the low-$E$ regime. According to the gap size, we can divide the phase diagram into three regions. The first one is the gapped zigzag ordered phase at low fields. The gap size is reduced with field, in concomitant with the suppression of the magnetic order with field as shown in Fig.~\ref{fig1}. Around the critical field, there is a narrow regime featuring {\new possibly} gapless continuous excitations. From our measurements, we estimate the range to be about 0.5~T. {\new By changing the way for the determination of  the gap, the evolution of the gap size with the magnetic field away from the critical regime may be different, but the gapless nature at 7.5 and 8~T remains to be the same.}  As we show above, the magnetic order is completely suppressed here, and the excitation continuum is consistent with gapless QSL, and is therefore labeled so. With further increasing field, the magnetic moments are forced to be aligned with the field. The gap reopens and the gap size increases monotonically with field. Since the saturation field is up to above 60~T~(Ref.~\onlinecite{PhysRevB.91.094422}), we label this phase as the gapped partially-polarized phase. We have measured the excitations with different $L$ values. It can be seen from Fig.~\ref{fig4} that while the gap size has some difference at different $L$s, the overall trend is similar. {\new Such a three-zone phase diagram with an intermediate QSL phase between the low-field zigzag ordered phase and a high-field partially polarized phase featuring two quantum critical points is different from those with only two phases---a low-field zigzag ordered phase and high-field QSL phase, divided by a quantum critical point\cite{PhysRevB.96.041405,PhysRevLett.119.037201}. The phase space for the QSL state is also significantly narrowed down. On the other hand, it is consistent with previous literature, which features a zigzag order, QSL, and possibly topologically trivial phase\cite{nature559_227,np17_915,wulferding2020magnon,PhysRevB.101.020414}.}

From our careful neutron scattering measurements on several batches of high-quality single crystals of \rucl on different spectrometers, we have obtained comprehensive excitation spectra as well as their evolutions under external magnetic field. These results allow us to answer the three important questions we raise above. First, the disordered phase near the critical field is indeed the QSL phase. This is evident from the fact that while the spin waves associated with the magnetic order are completely suppressed, the continuum associated with the fractional excitations of the QSL state survives. Second, the QSL phase is {\new possibly} gapless. We note that in Refs.~\onlinecite{nature559_227,doi:10.1126/science.aay5551}, a half-integer quantized plateau of the thermal Hall conductivity around the critical field was reported. This indicates the presence of a Kitaev QSL state with gapped bulk and gapless edge featuring Majorana fermions. On the other hand, another report on the thermal Hall conductivity did not observe the 1/2 plateau\cite{np17_915}. While this controversy remains to be solved, we believe that our direct measurements on the magnetic excitations by INS provide clear evidence that the excitations of the QSL are {\new very likely to be} gapless. Third, there is a small but finite intermediate QSL regime between the gapped zigzag order and gapped partially-polarized phase. By comparing the spectra of this field-induced gapless QSL phase [Fig.~\ref{fig3}(b)] with those of the zero-field paramagnetic phase above $T_{\rm N}$ [Fig.~\ref{fig3}(d)], we believe they are distinctive, although the latter phase may also feature fractional excitations\cite{np13_1079,PhysRevResearch.2.043015}.

To summarize, we have conducted INS experiments on \rucl single crystals to investigate the evolution of the magnetic excitations with an in-plane magnetic field. Our results show clearly that the spin-waves excitations around the $M$ point are suppressed in accordance with the suppression of the magnetic order. Near the critical field of $\sim$7.5~T, the spin waves disappear but the {\new possibly} gapless continuum around the $\mit\Gamma$ point is present, indicating the emergence of a pure gapless QSL phase. Based on the evolution of the gap with field, we obtain a three-zone phase diagram, which consists of a low-field gapped zigzag order phase, an intermediate-field gapless QSL, and a high-field gapped partially-polarized state. These results constitute as evidence that an intermediate in-plane magnetic field induces a pure QSL phase in \rucl.

The work was supported by National Key Projects for Research and Development of China with Grant No.~2021YFA1400400, the National Natural Science Foundation of China with Grant Nos.~11822405, 12074174, 12074175, 92165205, 11904170, 12004249, 12004251 and 12004191, Natural Science Foundation of Jiangsu Province with Grant Nos.~BK20180006, BK20190436 and BK20200738, Shanghai Sailing Program with Grant Nos.~ 20YF1430600 and 21YF1429200, and Fundamental Research Funds for the Central Universities. The experiment at FLEXX was carried out under proposal No.~17205993-CR using beamtime allocated in the HZB-CIAE collaboration on the scientific use of instruments. The experiments at SIKA were carried out under proposal No.~P5844 and No.~IC6741.


\begin{thebibliography}{85}%
\makeatletter
\providecommand \@ifxundefined [1]{%
 \@ifx{#1\undefined}
}%
\providecommand \@ifnum [1]{%
 \ifnum #1\expandafter \@firstoftwo
 \else \expandafter \@secondoftwo
 \fi
}%
\providecommand \@ifx [1]{%
 \ifx #1\expandafter \@firstoftwo
 \else \expandafter \@secondoftwo
 \fi
}%
\providecommand \natexlab [1]{#1}%
\providecommand \enquote  [1]{``#1''}%
\providecommand \bibnamefont  [1]{#1}%
\providecommand \bibfnamefont [1]{#1}%
\providecommand \citenamefont [1]{#1}%
\providecommand \href@noop [0]{\@secondoftwo}%
\providecommand \href [0]{\begingroup \@sanitize@url \@href}%
\providecommand \@href[1]{\@@startlink{#1}\@@href}%
\providecommand \@@href[1]{\endgroup#1\@@endlink}%
\providecommand \@sanitize@url [0]{\catcode `\\12\catcode `\$12\catcode
  `\&12\catcode `\#12\catcode `\^12\catcode `\_12\catcode `\%12\relax}%
\providecommand \@@startlink[1]{}%
\providecommand \@@endlink[0]{}%
\providecommand \url  [0]{\begingroup\@sanitize@url \@url }%
\providecommand \@url [1]{\endgroup\@href {#1}{\urlprefix }}%
\providecommand \urlprefix  [0]{URL }%
\providecommand \Eprint [0]{\href }%
\providecommand \doibase [0]{http://dx.doi.org/}%
\providecommand \selectlanguage [0]{\@gobble}%
\providecommand \bibinfo  [0]{\@secondoftwo}%
\providecommand \bibfield  [0]{\@secondoftwo}%
\providecommand \translation [1]{[#1]}%
\providecommand \BibitemOpen [0]{}%
\providecommand \bibitemStop [0]{}%
\providecommand \bibitemNoStop [0]{.\EOS\space}%
\providecommand \EOS [0]{\spacefactor3000\relax}%
\providecommand \BibitemShut  [1]{\csname bibitem#1\endcsname}%
\let\auto@bib@innerbib\@empty
\bibitem [{\citenamefont {Kitaev}(2006)}]{aop321_2}%
  \BibitemOpen
  \bibfield  {author} {\bibinfo {author} {\bibfnamefont {Alexei}\ \bibnamefont
  {Kitaev}},\ }\bibfield  {title} {\enquote {\bibinfo {title} {Anyons in an
  exactly solved model and beyond},}\ }\href@noop {} {\bibfield  {journal}
  {\bibinfo  {journal} {Ann. Phys.}\ }\textbf {\bibinfo {volume} {321}},\
  \bibinfo {pages} {2--111} (\bibinfo {year} {2006})}\BibitemShut {NoStop}%
\bibitem [{\citenamefont {Savary}\ and\ \citenamefont
  {Balents}(2017)}]{0034-4885-80-1-016502}%
  \BibitemOpen
  \bibfield  {author} {\bibinfo {author} {\bibfnamefont {Lucile}\ \bibnamefont
  {Savary}}\ and\ \bibinfo {author} {\bibfnamefont {Leon}\ \bibnamefont
  {Balents}},\ }\bibfield  {title} {\enquote {\bibinfo {title} {Quantum spin
  liquids: a review},}\ }\href
  {http://stacks.iop.org/0034-4885/80/i=1/a=016502} {\bibfield  {journal}
  {\bibinfo  {journal} {Rep. Pro. Phys.}\ }\textbf {\bibinfo {volume} {80}},\
  \bibinfo {pages} {016502} (\bibinfo {year} {2017})}\BibitemShut {NoStop}%
\bibitem [{\citenamefont {Takagi}\ \emph {et~al.}(2019)\citenamefont {Takagi},
  \citenamefont {Takayama}, \citenamefont {Jackeli}, \citenamefont
  {Khaliullin},\ and\ \citenamefont {Nagler}}]{nrp1_264}%
  \BibitemOpen
  \bibfield  {author} {\bibinfo {author} {\bibfnamefont {Hidenori}\
  \bibnamefont {Takagi}}, \bibinfo {author} {\bibfnamefont {Tomohiro}\
  \bibnamefont {Takayama}}, \bibinfo {author} {\bibfnamefont {George}\
  \bibnamefont {Jackeli}}, \bibinfo {author} {\bibfnamefont {Giniyat}\
  \bibnamefont {Khaliullin}}, \ and\ \bibinfo {author} {\bibfnamefont
  {Stephen~E.}\ \bibnamefont {Nagler}},\ }\bibfield  {title} {\enquote
  {\bibinfo {title} {{Concept and realization of Kitaev quantum spin
  liquids}},}\ }\href@noop {} {\bibfield  {journal} {\bibinfo  {journal} {Nat.
  Rev. Phys.}\ }\textbf {\bibinfo {volume} {1}},\ \bibinfo {pages} {264--280}
  (\bibinfo {year} {2019})}\BibitemShut {NoStop}%
\bibitem [{\citenamefont {Wen}\ \emph {et~al.}(2019)\citenamefont {Wen},
  \citenamefont {Yu}, \citenamefont {Li}, \citenamefont {Yu},\ and\
  \citenamefont {Li}}]{npjqm4_12}%
  \BibitemOpen
  \bibfield  {author} {\bibinfo {author} {\bibfnamefont {Jinsheng}\
  \bibnamefont {Wen}}, \bibinfo {author} {\bibfnamefont {Shun-Li}\ \bibnamefont
  {Yu}}, \bibinfo {author} {\bibfnamefont {Shiyan}\ \bibnamefont {Li}},
  \bibinfo {author} {\bibfnamefont {Weiqiang}\ \bibnamefont {Yu}}, \ and\
  \bibinfo {author} {\bibfnamefont {Jian-Xin}\ \bibnamefont {Li}},\ }\bibfield
  {title} {\enquote {\bibinfo {title} {Experimental identification of quantum
  spin liquids},}\ }\href@noop {} {\bibfield  {journal} {\bibinfo  {journal}
  {npj Quant. Mater.}\ }\textbf {\bibinfo {volume} {4}},\ \bibinfo {pages} {12}
  (\bibinfo {year} {2019})}\BibitemShut {NoStop}%
\bibitem [{\citenamefont {Anderson}(1973)}]{Anderson1973153}%
  \BibitemOpen
  \bibfield  {author} {\bibinfo {author} {\bibfnamefont {P.W.}\ \bibnamefont
  {Anderson}},\ }\bibfield  {title} {\enquote {\bibinfo {title} {{Resonating
  valence bonds: A new kind of insulator?}}}\ }\href@noop {} {\bibfield
  {journal} {\bibinfo  {journal} {Mater. Res. Bull.}\ }\textbf {\bibinfo
  {volume} {8}},\ \bibinfo {pages} {153--160} (\bibinfo {year}
  {1973})}\BibitemShut {NoStop}%
\bibitem [{\citenamefont {Balents}(2010)}]{nature464_199}%
  \BibitemOpen
  \bibfield  {author} {\bibinfo {author} {\bibfnamefont {Leon}\ \bibnamefont
  {Balents}},\ }\bibfield  {title} {\enquote {\bibinfo {title} {{Spin liquids
  in frustrated magnets}},}\ }\href {\doibase 10.1038/nature08917} {\bibfield
  {journal} {\bibinfo  {journal} {Nature}\ }\textbf {\bibinfo {volume} {464}},\
  \bibinfo {pages} {199--208} (\bibinfo {year} {2010})}\BibitemShut {NoStop}%
\bibitem [{\citenamefont {Sears}\ \emph {et~al.}(2015)\citenamefont {Sears},
  \citenamefont {Songvilay}, \citenamefont {Plumb}, \citenamefont {Clancy},
  \citenamefont {Qiu}, \citenamefont {Zhao}, \citenamefont {Parshall},\ and\
  \citenamefont {Kim}}]{PhysRevB.91.144420}%
  \BibitemOpen
  \bibfield  {author} {\bibinfo {author} {\bibfnamefont {J.~A.}\ \bibnamefont
  {Sears}}, \bibinfo {author} {\bibfnamefont {M.}~\bibnamefont {Songvilay}},
  \bibinfo {author} {\bibfnamefont {K.~W.}\ \bibnamefont {Plumb}}, \bibinfo
  {author} {\bibfnamefont {J.~P.}\ \bibnamefont {Clancy}}, \bibinfo {author}
  {\bibfnamefont {Y.}~\bibnamefont {Qiu}}, \bibinfo {author} {\bibfnamefont
  {Y.}~\bibnamefont {Zhao}}, \bibinfo {author} {\bibfnamefont {D.}~\bibnamefont
  {Parshall}}, \ and\ \bibinfo {author} {\bibfnamefont {Young-June}\
  \bibnamefont {Kim}},\ }\bibfield  {title} {\enquote {\bibinfo {title}
  {{Magnetic order in $\ensuremath{\alpha}-{\text{RuCl}}_{3}$: A
  honeycomb-lattice quantum magnet with strong spin-orbit coupling}},}\ }\href
  {\doibase 10.1103/PhysRevB.91.144420} {\bibfield  {journal} {\bibinfo
  {journal} {Phys. Rev. B}\ }\textbf {\bibinfo {volume} {91}},\ \bibinfo
  {pages} {144420} (\bibinfo {year} {2015})}\BibitemShut {NoStop}%
\bibitem [{\citenamefont {Kim}\ \emph {et~al.}(2015)\citenamefont {Kim},
  \citenamefont {V.}, \citenamefont {Catuneanu},\ and\ \citenamefont
  {Kee}}]{PhysRevB.91.241110}%
  \BibitemOpen
  \bibfield  {author} {\bibinfo {author} {\bibfnamefont {Heung-Sik}\
  \bibnamefont {Kim}}, \bibinfo {author} {\bibfnamefont {Vijay~Shankar}\
  \bibnamefont {V.}}, \bibinfo {author} {\bibfnamefont {Andrei}\ \bibnamefont
  {Catuneanu}}, \ and\ \bibinfo {author} {\bibfnamefont {Hae-Young}\
  \bibnamefont {Kee}},\ }\bibfield  {title} {\enquote {\bibinfo {title}
  {{Kitaev magnetism in honeycomb ${\text{RuCl}}_{3}$ with intermediate
  spin-orbit coupling}},}\ }\href {\doibase 10.1103/PhysRevB.91.241110}
  {\bibfield  {journal} {\bibinfo  {journal} {Phys. Rev. B}\ }\textbf {\bibinfo
  {volume} {91}},\ \bibinfo {pages} {241110} (\bibinfo {year}
  {2015})}\BibitemShut {NoStop}%
\bibitem [{\citenamefont {Johnson}\ \emph {et~al.}(2015)\citenamefont
  {Johnson}, \citenamefont {Williams}, \citenamefont {Haghighirad},
  \citenamefont {Singleton}, \citenamefont {Zapf}, \citenamefont {Manuel},
  \citenamefont {Mazin}, \citenamefont {Li}, \citenamefont {Jeschke},
  \citenamefont {Valent\'{\i}},\ and\ \citenamefont
  {Coldea}}]{PhysRevB.92.235119}%
  \BibitemOpen
  \bibfield  {author} {\bibinfo {author} {\bibfnamefont {R.~D.}\ \bibnamefont
  {Johnson}}, \bibinfo {author} {\bibfnamefont {S.~C.}\ \bibnamefont
  {Williams}}, \bibinfo {author} {\bibfnamefont {A.~A.}\ \bibnamefont
  {Haghighirad}}, \bibinfo {author} {\bibfnamefont {J.}~\bibnamefont
  {Singleton}}, \bibinfo {author} {\bibfnamefont {V.}~\bibnamefont {Zapf}},
  \bibinfo {author} {\bibfnamefont {P.}~\bibnamefont {Manuel}}, \bibinfo
  {author} {\bibfnamefont {I.~I.}\ \bibnamefont {Mazin}}, \bibinfo {author}
  {\bibfnamefont {Y.}~\bibnamefont {Li}}, \bibinfo {author} {\bibfnamefont
  {H.~O.}\ \bibnamefont {Jeschke}}, \bibinfo {author} {\bibfnamefont
  {R.}~\bibnamefont {Valent\'{\i}}}, \ and\ \bibinfo {author} {\bibfnamefont
  {R.}~\bibnamefont {Coldea}},\ }\bibfield  {title} {\enquote {\bibinfo {title}
  {{Monoclinic crystal structure of $\ensuremath{\alpha}-{\mathrm{RuCl}}_{3}$
  and the zigzag antiferromagnetic ground state}},}\ }\href {\doibase
  10.1103/PhysRevB.92.235119} {\bibfield  {journal} {\bibinfo  {journal} {Phys.
  Rev. B}\ }\textbf {\bibinfo {volume} {92}},\ \bibinfo {pages} {235119}
  (\bibinfo {year} {2015})}\BibitemShut {NoStop}%
\bibitem [{\citenamefont {Banerjee}\ \emph {et~al.}(2016)\citenamefont
  {Banerjee}, \citenamefont {Bridges}, \citenamefont {Yan}, \citenamefont
  {Aczel}, \citenamefont {Li}, \citenamefont {Stone}, \citenamefont {Granroth},
  \citenamefont {Lumsden}, \citenamefont {Yiu}, \citenamefont {Knolle},
  \citenamefont {Bhattacharjee}, \citenamefont {Kovrizhin}, \citenamefont
  {Moessner}, \citenamefont {Tennant}, \citenamefont {Mandrus},\ and\
  \citenamefont {Nagler}}]{nm15_733}%
  \BibitemOpen
  \bibfield  {author} {\bibinfo {author} {\bibfnamefont {A.}~\bibnamefont
  {Banerjee}}, \bibinfo {author} {\bibfnamefont {C.~A.}\ \bibnamefont
  {Bridges}}, \bibinfo {author} {\bibfnamefont {J.~Q.}\ \bibnamefont {Yan}},
  \bibinfo {author} {\bibfnamefont {A.~A.}\ \bibnamefont {Aczel}}, \bibinfo
  {author} {\bibfnamefont {L.}~\bibnamefont {Li}}, \bibinfo {author}
  {\bibfnamefont {M.~B.}\ \bibnamefont {Stone}}, \bibinfo {author}
  {\bibfnamefont {G.~E.}\ \bibnamefont {Granroth}}, \bibinfo {author}
  {\bibfnamefont {M.~D.}\ \bibnamefont {Lumsden}}, \bibinfo {author}
  {\bibfnamefont {Y.}~\bibnamefont {Yiu}}, \bibinfo {author} {\bibfnamefont
  {J.}~\bibnamefont {Knolle}}, \bibinfo {author} {\bibfnamefont
  {S.}~\bibnamefont {Bhattacharjee}}, \bibinfo {author} {\bibfnamefont {D.~L.}\
  \bibnamefont {Kovrizhin}}, \bibinfo {author} {\bibfnamefont {R.}~\bibnamefont
  {Moessner}}, \bibinfo {author} {\bibfnamefont {D.~A.}\ \bibnamefont
  {Tennant}}, \bibinfo {author} {\bibfnamefont {D.~G.}\ \bibnamefont
  {Mandrus}}, \ and\ \bibinfo {author} {\bibfnamefont {S.~E.}\ \bibnamefont
  {Nagler}},\ }\bibfield  {title} {\enquote {\bibinfo {title} {{Proximate
  Kitaev quantum spin liquid behaviour in a honeycomb magnet}},}\ }\href
  {\doibase 10.1038/nmat4604} {\bibfield  {journal} {\bibinfo  {journal} {Nat.
  Mater.}\ }\textbf {\bibinfo {volume} {15}},\ \bibinfo {pages} {733--740}
  (\bibinfo {year} {2016})}\BibitemShut {NoStop}%
\bibitem [{\citenamefont {Ritter}(2016)}]{Ritter_2016}%
  \BibitemOpen
  \bibfield  {author} {\bibinfo {author} {\bibfnamefont {C.}~\bibnamefont
  {Ritter}},\ }\bibfield  {title} {\enquote {\bibinfo {title} {{Zigzag type
  magnetic structure of the spin $J_{\rm eff}=1/2$ compound
  $\ensuremath{\alpha}\text{\ensuremath{-}}{\mathrm{RuCl}}_{3}$ as determined
  by neutron powder diffraction}},}\ }\href {\doibase
  10.1088/1742-6596/746/1/012060} {\bibfield  {journal} {\bibinfo  {journal}
  {J. Phys.: Conf. Ser.}\ }\textbf {\bibinfo {volume} {746}},\ \bibinfo {pages}
  {012060} (\bibinfo {year} {2016})}\BibitemShut {NoStop}%
\bibitem [{\citenamefont {Plumb}\ \emph {et~al.}(2014)\citenamefont {Plumb},
  \citenamefont {Clancy}, \citenamefont {Sandilands}, \citenamefont {Shankar},
  \citenamefont {Hu}, \citenamefont {Burch}, \citenamefont {Kee},\ and\
  \citenamefont {Kim}}]{PhysRevB.90.041112}%
  \BibitemOpen
  \bibfield  {author} {\bibinfo {author} {\bibfnamefont {K.~W.}\ \bibnamefont
  {Plumb}}, \bibinfo {author} {\bibfnamefont {J.~P.}\ \bibnamefont {Clancy}},
  \bibinfo {author} {\bibfnamefont {L.~J.}\ \bibnamefont {Sandilands}},
  \bibinfo {author} {\bibfnamefont {V.~Vijay}\ \bibnamefont {Shankar}},
  \bibinfo {author} {\bibfnamefont {Y.~F.}\ \bibnamefont {Hu}}, \bibinfo
  {author} {\bibfnamefont {K.~S.}\ \bibnamefont {Burch}}, \bibinfo {author}
  {\bibfnamefont {Hae-Young}\ \bibnamefont {Kee}}, \ and\ \bibinfo {author}
  {\bibfnamefont {Young-June}\ \bibnamefont {Kim}},\ }\bibfield  {title}
  {\enquote {\bibinfo {title} {{$\ensuremath{\alpha}-{\mathrm{RuCl}}_{3}$: A
  spin-orbit assisted Mott insulator on a honeycomb lattice}},}\ }\href
  {\doibase 10.1103/PhysRevB.90.041112} {\bibfield  {journal} {\bibinfo
  {journal} {Phys. Rev. B}\ }\textbf {\bibinfo {volume} {90}},\ \bibinfo
  {pages} {041112} (\bibinfo {year} {2014})}\BibitemShut {NoStop}%
\bibitem [{\citenamefont {Kim}\ and\ \citenamefont
  {Kee}(2016)}]{PhysRevB.93.155143}%
  \BibitemOpen
  \bibfield  {author} {\bibinfo {author} {\bibfnamefont {Heung-Sik}\
  \bibnamefont {Kim}}\ and\ \bibinfo {author} {\bibfnamefont {Hae-Young}\
  \bibnamefont {Kee}},\ }\bibfield  {title} {\enquote {\bibinfo {title}
  {{Crystal structure and magnetism in
  $\ensuremath{\alpha}-{\mathrm{RuCl}}_{3}$: An \textit{ab initio} study}},}\
  }\href {\doibase 10.1103/PhysRevB.93.155143} {\bibfield  {journal} {\bibinfo
  {journal} {Phys. Rev. B}\ }\textbf {\bibinfo {volume} {93}},\ \bibinfo
  {pages} {155143} (\bibinfo {year} {2016})}\BibitemShut {NoStop}%
\bibitem [{\citenamefont {Yadav}\ \emph {et~al.}(2016)\citenamefont {Yadav},
  \citenamefont {Bogdanov}, \citenamefont {Katukuri}, \citenamefont
  {Nishimoto}, \citenamefont {van~den Brink},\ and\ \citenamefont
  {Hozoi}}]{sr6_37925}%
  \BibitemOpen
  \bibfield  {author} {\bibinfo {author} {\bibfnamefont {Ravi}\ \bibnamefont
  {Yadav}}, \bibinfo {author} {\bibfnamefont {Nikolay~A.}\ \bibnamefont
  {Bogdanov}}, \bibinfo {author} {\bibfnamefont {Vamshi~M.}\ \bibnamefont
  {Katukuri}}, \bibinfo {author} {\bibfnamefont {Satoshi}\ \bibnamefont
  {Nishimoto}}, \bibinfo {author} {\bibfnamefont {Jeroen}\ \bibnamefont
  {van~den Brink}}, \ and\ \bibinfo {author} {\bibfnamefont {Liviu}\
  \bibnamefont {Hozoi}},\ }\bibfield  {title} {\enquote {\bibinfo {title}
  {{Kitaev exchange and field-induced quantum spin-liquid states in honeycomb
  $\ensuremath{\alpha}\text{\ensuremath{-}}{\mathrm{RuCl}}_{3}$}},}\
  }\href@noop {} {\bibfield  {journal} {\bibinfo  {journal} {Sci. Rep.}\
  }\textbf {\bibinfo {volume} {6}},\ \bibinfo {pages} {37925} (\bibinfo {year}
  {2016})}\BibitemShut {NoStop}%
\bibitem [{\citenamefont {Ran}\ \emph {et~al.}(2017)\citenamefont {Ran},
  \citenamefont {Wang}, \citenamefont {Wang}, \citenamefont {Dong},
  \citenamefont {Ren}, \citenamefont {Bao}, \citenamefont {Li}, \citenamefont
  {Ma}, \citenamefont {Gan}, \citenamefont {Zhang}, \citenamefont {Park},
  \citenamefont {Deng}, \citenamefont {Danilkin}, \citenamefont {Yu},
  \citenamefont {Li},\ and\ \citenamefont {Wen}}]{PhysRevLett.118.107203}%
  \BibitemOpen
  \bibfield  {author} {\bibinfo {author} {\bibfnamefont {Kejing}\ \bibnamefont
  {Ran}}, \bibinfo {author} {\bibfnamefont {Jinghui}\ \bibnamefont {Wang}},
  \bibinfo {author} {\bibfnamefont {Wei}\ \bibnamefont {Wang}}, \bibinfo
  {author} {\bibfnamefont {Zhao-Yang}\ \bibnamefont {Dong}}, \bibinfo {author}
  {\bibfnamefont {Xiao}\ \bibnamefont {Ren}}, \bibinfo {author} {\bibfnamefont
  {Song}\ \bibnamefont {Bao}}, \bibinfo {author} {\bibfnamefont {Shichao}\
  \bibnamefont {Li}}, \bibinfo {author} {\bibfnamefont {Zhen}\ \bibnamefont
  {Ma}}, \bibinfo {author} {\bibfnamefont {Yuan}\ \bibnamefont {Gan}}, \bibinfo
  {author} {\bibfnamefont {Youtian}\ \bibnamefont {Zhang}}, \bibinfo {author}
  {\bibfnamefont {J.~T.}\ \bibnamefont {Park}}, \bibinfo {author}
  {\bibfnamefont {Guochu}\ \bibnamefont {Deng}}, \bibinfo {author}
  {\bibfnamefont {S.}~\bibnamefont {Danilkin}}, \bibinfo {author}
  {\bibfnamefont {Shun-Li}\ \bibnamefont {Yu}}, \bibinfo {author}
  {\bibfnamefont {Jian-Xin}\ \bibnamefont {Li}}, \ and\ \bibinfo {author}
  {\bibfnamefont {Jinsheng}\ \bibnamefont {Wen}},\ }\bibfield  {title}
  {\enquote {\bibinfo {title} {{Spin-Wave Excitations Evidencing the Kitaev
  Interaction in Single Crystalline
  $\ensuremath{\alpha}\text{\ensuremath{-}}{\mathrm{RuCl}}_{3}$}},}\ }\href
  {\doibase 10.1103/PhysRevLett.118.107203} {\bibfield  {journal} {\bibinfo
  {journal} {Phys. Rev. Lett.}\ }\textbf {\bibinfo {volume} {118}},\ \bibinfo
  {pages} {107203} (\bibinfo {year} {2017})}\BibitemShut {NoStop}%
\bibitem [{\citenamefont {Wang}\ \emph
  {et~al.}(2017{\natexlab{a}})\citenamefont {Wang}, \citenamefont {Dong},
  \citenamefont {Yu},\ and\ \citenamefont {Li}}]{PhysRevB.96.115103}%
  \BibitemOpen
  \bibfield  {author} {\bibinfo {author} {\bibfnamefont {Wei}\ \bibnamefont
  {Wang}}, \bibinfo {author} {\bibfnamefont {Zhao-Yang}\ \bibnamefont {Dong}},
  \bibinfo {author} {\bibfnamefont {Shun-Li}\ \bibnamefont {Yu}}, \ and\
  \bibinfo {author} {\bibfnamefont {Jian-Xin}\ \bibnamefont {Li}},\ }\bibfield
  {title} {\enquote {\bibinfo {title} {{Theoretical investigation of magnetic
  dynamics in $\ensuremath{\alpha}\ensuremath{-}{\mathrm{RuCl}}_{3}$}},}\
  }\href {\doibase 10.1103/PhysRevB.96.115103} {\bibfield  {journal} {\bibinfo
  {journal} {Phys. Rev. B}\ }\textbf {\bibinfo {volume} {96}},\ \bibinfo
  {pages} {115103} (\bibinfo {year} {2017}{\natexlab{a}})}\BibitemShut
  {NoStop}%
\bibitem [{\citenamefont {Sears}\ \emph {et~al.}(2020)\citenamefont {Sears},
  \citenamefont {Chern}, \citenamefont {Kim}, \citenamefont {Bereciartua},
  \citenamefont {Francoual}, \citenamefont {Kim},\ and\ \citenamefont
  {Kim}}]{np16_837}%
  \BibitemOpen
  \bibfield  {author} {\bibinfo {author} {\bibfnamefont {Jennifer~A.}\
  \bibnamefont {Sears}}, \bibinfo {author} {\bibfnamefont {Li~Ern}\
  \bibnamefont {Chern}}, \bibinfo {author} {\bibfnamefont {Subin}\ \bibnamefont
  {Kim}}, \bibinfo {author} {\bibfnamefont {Pablo~J.}\ \bibnamefont
  {Bereciartua}}, \bibinfo {author} {\bibfnamefont {Sonia}\ \bibnamefont
  {Francoual}}, \bibinfo {author} {\bibfnamefont {Yong~Baek}\ \bibnamefont
  {Kim}}, \ and\ \bibinfo {author} {\bibfnamefont {Young-June}\ \bibnamefont
  {Kim}},\ }\bibfield  {title} {\enquote {\bibinfo {title} {{Ferromagnetic
  Kitaev interaction and the origin of large magnetic anisotropy in
  $\ensuremath{\alpha}\text{\ensuremath{-}}{\mathrm{RuCl}}_{3}$}},}\
  }\href@noop {} {\bibfield  {journal} {\bibinfo  {journal} {Nat. Phys.}\
  }\textbf {\bibinfo {volume} {16}},\ \bibinfo {pages} {837--840} (\bibinfo
  {year} {2020})}\BibitemShut {NoStop}%
\bibitem [{\citenamefont {Sandilands}\ \emph {et~al.}(2015)\citenamefont
  {Sandilands}, \citenamefont {Tian}, \citenamefont {Plumb}, \citenamefont
  {Kim},\ and\ \citenamefont {Burch}}]{PhysRevLett.114.147201}%
  \BibitemOpen
  \bibfield  {author} {\bibinfo {author} {\bibfnamefont {Luke~J.}\ \bibnamefont
  {Sandilands}}, \bibinfo {author} {\bibfnamefont {Yao}\ \bibnamefont {Tian}},
  \bibinfo {author} {\bibfnamefont {Kemp~W.}\ \bibnamefont {Plumb}}, \bibinfo
  {author} {\bibfnamefont {Young-June}\ \bibnamefont {Kim}}, \ and\ \bibinfo
  {author} {\bibfnamefont {Kenneth~S.}\ \bibnamefont {Burch}},\ }\bibfield
  {title} {\enquote {\bibinfo {title} {{Scattering Continuum and Possible
  Fractionalized Excitations in
  $\ensuremath{\alpha}\text{\ensuremath{-}}{\mathrm{RuCl}}_{3}$}},}\ }\href
  {\doibase 10.1103/PhysRevLett.114.147201} {\bibfield  {journal} {\bibinfo
  {journal} {Phys. Rev. Lett.}\ }\textbf {\bibinfo {volume} {114}},\ \bibinfo
  {pages} {147201} (\bibinfo {year} {2015})}\BibitemShut {NoStop}%
\bibitem [{\citenamefont {Nasu}\ \emph {et~al.}(2016)\citenamefont {Nasu},
  \citenamefont {Knolle}, \citenamefont {Kovrizhin}, \citenamefont {Motome},\
  and\ \citenamefont {Moessner}}]{np12_912}%
  \BibitemOpen
  \bibfield  {author} {\bibinfo {author} {\bibfnamefont {J.}~\bibnamefont
  {Nasu}}, \bibinfo {author} {\bibfnamefont {J.}~\bibnamefont {Knolle}},
  \bibinfo {author} {\bibfnamefont {D.~L.}\ \bibnamefont {Kovrizhin}}, \bibinfo
  {author} {\bibfnamefont {Y.}~\bibnamefont {Motome}}, \ and\ \bibinfo {author}
  {\bibfnamefont {R.}~\bibnamefont {Moessner}},\ }\bibfield  {title} {\enquote
  {\bibinfo {title} {Fermionic response from fractionalization in an insulating
  two-dimensional magnet},}\ }\href@noop {} {\bibfield  {journal} {\bibinfo
  {journal} {Nat. Phys.}\ }\textbf {\bibinfo {volume} {12}},\ \bibinfo {pages}
  {912--915} (\bibinfo {year} {2016})}\BibitemShut {NoStop}%
\bibitem [{\citenamefont {Do}\ \emph {et~al.}(2017)\citenamefont {Do},
  \citenamefont {Park}, \citenamefont {Yoshitake}, \citenamefont {Nasu},
  \citenamefont {Motome}, \citenamefont {Kwon}, \citenamefont {Adroja},
  \citenamefont {Voneshen}, \citenamefont {Kim}, \citenamefont {Jang},
  \citenamefont {Park}, \citenamefont {Choi},\ and\ \citenamefont
  {Ji}}]{np13_1079}%
  \BibitemOpen
  \bibfield  {author} {\bibinfo {author} {\bibfnamefont {Seung-Hwan}\
  \bibnamefont {Do}}, \bibinfo {author} {\bibfnamefont {Sang-Youn}\
  \bibnamefont {Park}}, \bibinfo {author} {\bibfnamefont {Junki}\ \bibnamefont
  {Yoshitake}}, \bibinfo {author} {\bibfnamefont {Joji}\ \bibnamefont {Nasu}},
  \bibinfo {author} {\bibfnamefont {Yukitoshi}\ \bibnamefont {Motome}},
  \bibinfo {author} {\bibfnamefont {Yong~Seung}\ \bibnamefont {Kwon}}, \bibinfo
  {author} {\bibfnamefont {D.~T.}\ \bibnamefont {Adroja}}, \bibinfo {author}
  {\bibfnamefont {D.~J.}\ \bibnamefont {Voneshen}}, \bibinfo {author}
  {\bibfnamefont {Kyoo}\ \bibnamefont {Kim}}, \bibinfo {author} {\bibfnamefont
  {T.~H.}\ \bibnamefont {Jang}}, \bibinfo {author} {\bibfnamefont {J.~H.}\
  \bibnamefont {Park}}, \bibinfo {author} {\bibfnamefont {Kwang-Yong}\
  \bibnamefont {Choi}}, \ and\ \bibinfo {author} {\bibfnamefont {Sungdae}\
  \bibnamefont {Ji}},\ }\bibfield  {title} {\enquote {\bibinfo {title}
  {{Majorana fermions in the Kitaev quantum spin system
  $\ensuremath{\alpha}\text{\ensuremath{-}}{\mathrm{RuCl}}_{3}$}},}\
  }\href@noop {} {\bibfield  {journal} {\bibinfo  {journal} {Nat. Phys.}\
  }\textbf {\bibinfo {volume} {13}},\ \bibinfo {pages} {1079} (\bibinfo {year}
  {2017})}\BibitemShut {NoStop}%
\bibitem [{\citenamefont {Hou}\ \emph {et~al.}(2017)\citenamefont {Hou},
  \citenamefont {Xiang},\ and\ \citenamefont {Gong}}]{PhysRevB.96.054410}%
  \BibitemOpen
  \bibfield  {author} {\bibinfo {author} {\bibfnamefont {Y.~S.}\ \bibnamefont
  {Hou}}, \bibinfo {author} {\bibfnamefont {H.~J.}\ \bibnamefont {Xiang}}, \
  and\ \bibinfo {author} {\bibfnamefont {X.~G.}\ \bibnamefont {Gong}},\
  }\bibfield  {title} {\enquote {\bibinfo {title} {{Unveiling magnetic
  interactions of ruthenium trichloride via constraining direction of orbital
  moments: Potential routes to realize a quantum spin liquid}},}\ }\href
  {\doibase 10.1103/PhysRevB.96.054410} {\bibfield  {journal} {\bibinfo
  {journal} {Phys. Rev. B}\ }\textbf {\bibinfo {volume} {96}},\ \bibinfo
  {pages} {054410} (\bibinfo {year} {2017})}\BibitemShut {NoStop}%
\bibitem [{\citenamefont {Lampen-Kelley}\ \emph {et~al.}(2018)\citenamefont
  {Lampen-Kelley}, \citenamefont {Rachel}, \citenamefont {Reuther},
  \citenamefont {Yan}, \citenamefont {Banerjee}, \citenamefont {Bridges},
  \citenamefont {Cao}, \citenamefont {Nagler},\ and\ \citenamefont
  {Mandrus}}]{PhysRevB.98.100403}%
  \BibitemOpen
  \bibfield  {author} {\bibinfo {author} {\bibfnamefont {P.}~\bibnamefont
  {Lampen-Kelley}}, \bibinfo {author} {\bibfnamefont {S.}~\bibnamefont
  {Rachel}}, \bibinfo {author} {\bibfnamefont {J.}~\bibnamefont {Reuther}},
  \bibinfo {author} {\bibfnamefont {J.-Q.}\ \bibnamefont {Yan}}, \bibinfo
  {author} {\bibfnamefont {A.}~\bibnamefont {Banerjee}}, \bibinfo {author}
  {\bibfnamefont {C.~A.}\ \bibnamefont {Bridges}}, \bibinfo {author}
  {\bibfnamefont {H.~B.}\ \bibnamefont {Cao}}, \bibinfo {author} {\bibfnamefont
  {S.~E.}\ \bibnamefont {Nagler}}, \ and\ \bibinfo {author} {\bibfnamefont
  {D.}~\bibnamefont {Mandrus}},\ }\bibfield  {title} {\enquote {\bibinfo
  {title} {{Anisotropic susceptibilities in the honeycomb Kitaev system
  $\ensuremath{\alpha}\ensuremath{-}{\mathrm{RuCl}}_{3}$}},}\ }\href {\doibase
  10.1103/PhysRevB.98.100403} {\bibfield  {journal} {\bibinfo  {journal} {Phys.
  Rev. B}\ }\textbf {\bibinfo {volume} {98}},\ \bibinfo {pages} {100403}
  (\bibinfo {year} {2018})}\BibitemShut {NoStop}%
\bibitem [{\citenamefont {Eichstaedt}\ \emph {et~al.}(2019)\citenamefont
  {Eichstaedt}, \citenamefont {Zhang}, \citenamefont {Laurell}, \citenamefont
  {Okamoto}, \citenamefont {Eguiluz},\ and\ \citenamefont
  {Berlijn}}]{PhysRevB.100.075110}%
  \BibitemOpen
  \bibfield  {author} {\bibinfo {author} {\bibfnamefont {Casey}\ \bibnamefont
  {Eichstaedt}}, \bibinfo {author} {\bibfnamefont {Yi}~\bibnamefont {Zhang}},
  \bibinfo {author} {\bibfnamefont {Pontus}\ \bibnamefont {Laurell}}, \bibinfo
  {author} {\bibfnamefont {Satoshi}\ \bibnamefont {Okamoto}}, \bibinfo {author}
  {\bibfnamefont {Adolfo~G.}\ \bibnamefont {Eguiluz}}, \ and\ \bibinfo {author}
  {\bibfnamefont {Tom}\ \bibnamefont {Berlijn}},\ }\bibfield  {title} {\enquote
  {\bibinfo {title} {{Deriving models for the Kitaev spin-liquid candidate
  material $\ensuremath{\alpha}\text{\ensuremath{-}}{\mathrm{RuCl}}_{3}$ from
  first principles}},}\ }\href {\doibase 10.1103/PhysRevB.100.075110}
  {\bibfield  {journal} {\bibinfo  {journal} {Phys. Rev. B}\ }\textbf {\bibinfo
  {volume} {100}},\ \bibinfo {pages} {075110} (\bibinfo {year}
  {2019})}\BibitemShut {NoStop}%
\bibitem [{\citenamefont {Chaloupka}\ \emph {et~al.}(2010)\citenamefont
  {Chaloupka}, \citenamefont {Jackeli},\ and\ \citenamefont
  {Khaliullin}}]{prl105_027204}%
  \BibitemOpen
  \bibfield  {author} {\bibinfo {author} {\bibfnamefont {Ji\v{r}\'{i}}\
  \bibnamefont {Chaloupka}}, \bibinfo {author} {\bibfnamefont {George}\
  \bibnamefont {Jackeli}}, \ and\ \bibinfo {author} {\bibfnamefont {Giniyat}\
  \bibnamefont {Khaliullin}},\ }\bibfield  {title} {\enquote {\bibinfo {title}
  {{Kitaev-Heisenberg Model on a Honeycomb Lattice: Possible Exotic Phases in
  Iridium Oxides ${A}_{2}{\mathrm{IrO}}_{3}$}},}\ }\href@noop {} {\bibfield
  {journal} {\bibinfo  {journal} {Phys. Rev. Lett.}\ }\textbf {\bibinfo
  {volume} {105}},\ \bibinfo {pages} {027204} (\bibinfo {year}
  {2010})}\BibitemShut {NoStop}%
\bibitem [{\citenamefont {Rau}\ \emph {et~al.}(2014)\citenamefont {Rau},
  \citenamefont {Lee},\ and\ \citenamefont {Kee}}]{PhysRevLett.112.077204}%
  \BibitemOpen
  \bibfield  {author} {\bibinfo {author} {\bibfnamefont {Jeffrey~G.}\
  \bibnamefont {Rau}}, \bibinfo {author} {\bibfnamefont {Eric Kin-Ho}\
  \bibnamefont {Lee}}, \ and\ \bibinfo {author} {\bibfnamefont {Hae-Young}\
  \bibnamefont {Kee}},\ }\bibfield  {title} {\enquote {\bibinfo {title}
  {{Generic Spin Model for the Honeycomb Iridates beyond the Kitaev Limit}},}\
  }\href {\doibase 10.1103/PhysRevLett.112.077204} {\bibfield  {journal}
  {\bibinfo  {journal} {Phys. Rev. Lett.}\ }\textbf {\bibinfo {volume} {112}},\
  \bibinfo {pages} {077204} (\bibinfo {year} {2014})}\BibitemShut {NoStop}%
\bibitem [{\citenamefont {Chaloupka}\ and\ \citenamefont
  {Khaliullin}(2015)}]{PhysRevB.92.024413}%
  \BibitemOpen
  \bibfield  {author} {\bibinfo {author} {\bibfnamefont {Ji\v{r}\'{i}}\
  \bibnamefont {Chaloupka}}\ and\ \bibinfo {author} {\bibfnamefont {Giniyat}\
  \bibnamefont {Khaliullin}},\ }\bibfield  {title} {\enquote {\bibinfo {title}
  {{Hidden symmetries of the extended Kitaev-Heisenberg model: Implications for
  the honeycomb-lattice iridates ${A}_{2}{\mathrm{IrO}}_{3}$}},}\ }\href
  {\doibase 10.1103/PhysRevB.92.024413} {\bibfield  {journal} {\bibinfo
  {journal} {Phys. Rev. B}\ }\textbf {\bibinfo {volume} {92}},\ \bibinfo
  {pages} {024413} (\bibinfo {year} {2015})}\BibitemShut {NoStop}%
\bibitem [{\citenamefont {Katukuri}\ \emph {et~al.}(2014)\citenamefont
  {Katukuri}, \citenamefont {Nishimoto}, \citenamefont {Yushankhai},
  \citenamefont {Stoyanova}, \citenamefont {Kandpal}, \citenamefont {Choi},
  \citenamefont {Coldea}, \citenamefont {Rousochatzakis}, \citenamefont
  {Hozoi},\ and\ \citenamefont {van~den Brink}}]{1367-2630-16-1-013056}%
  \BibitemOpen
  \bibfield  {author} {\bibinfo {author} {\bibfnamefont {Vamshi~M}\
  \bibnamefont {Katukuri}}, \bibinfo {author} {\bibfnamefont {S}~\bibnamefont
  {Nishimoto}}, \bibinfo {author} {\bibfnamefont {V}~\bibnamefont
  {Yushankhai}}, \bibinfo {author} {\bibfnamefont {A}~\bibnamefont
  {Stoyanova}}, \bibinfo {author} {\bibfnamefont {H}~\bibnamefont {Kandpal}},
  \bibinfo {author} {\bibfnamefont {Sungkyun}\ \bibnamefont {Choi}}, \bibinfo
  {author} {\bibfnamefont {R}~\bibnamefont {Coldea}}, \bibinfo {author}
  {\bibfnamefont {I}~\bibnamefont {Rousochatzakis}}, \bibinfo {author}
  {\bibfnamefont {L}~\bibnamefont {Hozoi}}, \ and\ \bibinfo {author}
  {\bibfnamefont {Jeroen}\ \bibnamefont {van~den Brink}},\ }\bibfield  {title}
  {\enquote {\bibinfo {title} {{Kitaev interactions between $j$ = 1/2 moments
  in honeycomb Na$_2$IrO$_3$ are large and ferromagnetic: insights from ab
  initio quantum chemistry calculations}},}\ }\href
  {http://stacks.iop.org/1367-2630/16/i=1/a=013056} {\bibfield  {journal}
  {\bibinfo  {journal} {New J. Phys.}\ }\textbf {\bibinfo {volume} {16}},\
  \bibinfo {pages} {013056} (\bibinfo {year} {2014})}\BibitemShut {NoStop}%
\bibitem [{\citenamefont {Yamaji}\ \emph {et~al.}(2014)\citenamefont {Yamaji},
  \citenamefont {Nomura}, \citenamefont {Kurita}, \citenamefont {Arita},\ and\
  \citenamefont {Imada}}]{PhysRevLett.113.107201}%
  \BibitemOpen
  \bibfield  {author} {\bibinfo {author} {\bibfnamefont {Youhei}\ \bibnamefont
  {Yamaji}}, \bibinfo {author} {\bibfnamefont {Yusuke}\ \bibnamefont {Nomura}},
  \bibinfo {author} {\bibfnamefont {Moyuru}\ \bibnamefont {Kurita}}, \bibinfo
  {author} {\bibfnamefont {Ryotaro}\ \bibnamefont {Arita}}, \ and\ \bibinfo
  {author} {\bibfnamefont {Masatoshi}\ \bibnamefont {Imada}},\ }\bibfield
  {title} {\enquote {\bibinfo {title} {{First-Principles Study of the
  Honeycomb-Lattice Iridates ${\mathrm{Na}}_{2}{\mathrm{IrO}}_{3}$ in the
  Presence of Strong Spin-Orbit Interaction and Electron Correlations}},}\
  }\href {\doibase 10.1103/PhysRevLett.113.107201} {\bibfield  {journal}
  {\bibinfo  {journal} {Phys. Rev. Lett.}\ }\textbf {\bibinfo {volume} {113}},\
  \bibinfo {pages} {107201} (\bibinfo {year} {2014})}\BibitemShut {NoStop}%
\bibitem [{\citenamefont {Hwan~Chun}\ \emph {et~al.}(2015)\citenamefont
  {Hwan~Chun}, \citenamefont {Kim}, \citenamefont {Kim}, \citenamefont {Zheng},
  \citenamefont {Stoumpos}, \citenamefont {Malliakas}, \citenamefont
  {Mitchell}, \citenamefont {Mehlawat}, \citenamefont {Singh}, \citenamefont
  {Choi}, \citenamefont {Gog}, \citenamefont {Al-Zein}, \citenamefont {Sala},
  \citenamefont {Krisch}, \citenamefont {Chaloupka}, \citenamefont {Jackeli},
  \citenamefont {Khaliullin},\ and\ \citenamefont {Kim}}]{np11_462}%
  \BibitemOpen
  \bibfield  {author} {\bibinfo {author} {\bibfnamefont {Sae}\ \bibnamefont
  {Hwan~Chun}}, \bibinfo {author} {\bibfnamefont {Jong-Woo}\ \bibnamefont
  {Kim}}, \bibinfo {author} {\bibfnamefont {Jungho}\ \bibnamefont {Kim}},
  \bibinfo {author} {\bibfnamefont {H.}~\bibnamefont {Zheng}}, \bibinfo
  {author} {\bibfnamefont {Constantinos~C.}\ \bibnamefont {Stoumpos}}, \bibinfo
  {author} {\bibfnamefont {C.~D.}\ \bibnamefont {Malliakas}}, \bibinfo {author}
  {\bibfnamefont {J.~F.}\ \bibnamefont {Mitchell}}, \bibinfo {author}
  {\bibfnamefont {Kavita}\ \bibnamefont {Mehlawat}}, \bibinfo {author}
  {\bibfnamefont {Yogesh}\ \bibnamefont {Singh}}, \bibinfo {author}
  {\bibfnamefont {Y.}~\bibnamefont {Choi}}, \bibinfo {author} {\bibfnamefont
  {T.}~\bibnamefont {Gog}}, \bibinfo {author} {\bibfnamefont {A.}~\bibnamefont
  {Al-Zein}}, \bibinfo {author} {\bibfnamefont {M.~Moretti}\ \bibnamefont
  {Sala}}, \bibinfo {author} {\bibfnamefont {M.}~\bibnamefont {Krisch}},
  \bibinfo {author} {\bibfnamefont {J.}~\bibnamefont {Chaloupka}}, \bibinfo
  {author} {\bibfnamefont {G.}~\bibnamefont {Jackeli}}, \bibinfo {author}
  {\bibfnamefont {G.}~\bibnamefont {Khaliullin}}, \ and\ \bibinfo {author}
  {\bibfnamefont {B.~J.}\ \bibnamefont {Kim}},\ }\bibfield  {title} {\enquote
  {\bibinfo {title} {{Direct evidence for dominant bond-directional
  interactions in a honeycomb lattice iridate Na$_2$IrO$_3$}},}\ }\href@noop {}
  {\bibfield  {journal} {\bibinfo  {journal} {Nat. Phys.}\ }\textbf {\bibinfo
  {volume} {11}},\ \bibinfo {pages} {462--466} (\bibinfo {year}
  {2015})}\BibitemShut {NoStop}%
\bibitem [{\citenamefont {Jackeli}\ and\ \citenamefont
  {Khaliullin}(2009)}]{prl102_017205}%
  \BibitemOpen
  \bibfield  {author} {\bibinfo {author} {\bibfnamefont {G.}~\bibnamefont
  {Jackeli}}\ and\ \bibinfo {author} {\bibfnamefont {G.}~\bibnamefont
  {Khaliullin}},\ }\bibfield  {title} {\enquote {\bibinfo {title} {{Mott
  Insulators in the Strong Spin-Orbit Coupling Limit: From Heisenberg to a
  Quantum Compass and Kitaev Models}},}\ }\href@noop {} {\bibfield  {journal}
  {\bibinfo  {journal} {Phys. Rev. Lett.}\ }\textbf {\bibinfo {volume} {102}},\
  \bibinfo {pages} {017205} (\bibinfo {year} {2009})}\BibitemShut {NoStop}%
\bibitem [{\citenamefont {Winter}\ \emph {et~al.}(2017)\citenamefont {Winter},
  \citenamefont {Tsirlin}, \citenamefont {Daghofer}, \citenamefont {van~den
  Brink}, \citenamefont {Singh}, \citenamefont {Gegenwart},\ and\ \citenamefont
  {Valent{\'{\i}}}}]{0953-8984-29-49-493002}%
  \BibitemOpen
  \bibfield  {author} {\bibinfo {author} {\bibfnamefont {Stephen~M}\
  \bibnamefont {Winter}}, \bibinfo {author} {\bibfnamefont {Alexander~A}\
  \bibnamefont {Tsirlin}}, \bibinfo {author} {\bibfnamefont {Maria}\
  \bibnamefont {Daghofer}}, \bibinfo {author} {\bibfnamefont {Jeroen}\
  \bibnamefont {van~den Brink}}, \bibinfo {author} {\bibfnamefont {Yogesh}\
  \bibnamefont {Singh}}, \bibinfo {author} {\bibfnamefont {Philipp}\
  \bibnamefont {Gegenwart}}, \ and\ \bibinfo {author} {\bibfnamefont {Roser}\
  \bibnamefont {Valent{\'{\i}}}},\ }\bibfield  {title} {\enquote {\bibinfo
  {title} {{Models and materials for generalized Kitaev magnetism}},}\ }\href
  {http://stacks.iop.org/0953-8984/29/i=49/a=493002} {\bibfield  {journal}
  {\bibinfo  {journal} {J. Phys. Conden. Matter}\ }\textbf {\bibinfo {volume}
  {29}},\ \bibinfo {pages} {493002} (\bibinfo {year} {2017})}\BibitemShut
  {NoStop}%
\bibitem [{\citenamefont {Winter}\ \emph {et~al.}(2016)\citenamefont {Winter},
  \citenamefont {Li}, \citenamefont {Jeschke},\ and\ \citenamefont
  {Valent\'{\i}}}]{PhysRevB.93.214431}%
  \BibitemOpen
  \bibfield  {author} {\bibinfo {author} {\bibfnamefont {Stephen~M.}\
  \bibnamefont {Winter}}, \bibinfo {author} {\bibfnamefont {Ying}\ \bibnamefont
  {Li}}, \bibinfo {author} {\bibfnamefont {Harald~O.}\ \bibnamefont {Jeschke}},
  \ and\ \bibinfo {author} {\bibfnamefont {Roser}\ \bibnamefont
  {Valent\'{\i}}},\ }\bibfield  {title} {\enquote {\bibinfo {title}
  {{Challenges in design of Kitaev materials: Magnetic interactions from
  competing energy scales}},}\ }\href {\doibase 10.1103/PhysRevB.93.214431}
  {\bibfield  {journal} {\bibinfo  {journal} {Phys. Rev. B}\ }\textbf {\bibinfo
  {volume} {93}},\ \bibinfo {pages} {214431} (\bibinfo {year}
  {2016})}\BibitemShut {NoStop}%
\bibitem [{\citenamefont {Ronquillo}\ \emph {et~al.}(2019)\citenamefont
  {Ronquillo}, \citenamefont {Vengal},\ and\ \citenamefont
  {Trivedi}}]{PhysRevB.99.140413}%
  \BibitemOpen
  \bibfield  {author} {\bibinfo {author} {\bibfnamefont {David~C.}\
  \bibnamefont {Ronquillo}}, \bibinfo {author} {\bibfnamefont {Adu}\
  \bibnamefont {Vengal}}, \ and\ \bibinfo {author} {\bibfnamefont {Nandini}\
  \bibnamefont {Trivedi}},\ }\bibfield  {title} {\enquote {\bibinfo {title}
  {{Signatures of magnetic-field-driven quantum phase transitions in the
  entanglement entropy and spin dynamics of the Kitaev honeycomb model}},}\
  }\href {\doibase 10.1103/PhysRevB.99.140413} {\bibfield  {journal} {\bibinfo
  {journal} {Phys. Rev. B}\ }\textbf {\bibinfo {volume} {99}},\ \bibinfo
  {pages} {140413} (\bibinfo {year} {2019})}\BibitemShut {NoStop}%
\bibitem [{\citenamefont {Janssen}\ \emph {et~al.}(2017)\citenamefont
  {Janssen}, \citenamefont {Andrade},\ and\ \citenamefont
  {Vojta}}]{PhysRevB.96.064430}%
  \BibitemOpen
  \bibfield  {author} {\bibinfo {author} {\bibfnamefont {Lukas}\ \bibnamefont
  {Janssen}}, \bibinfo {author} {\bibfnamefont {Eric~C.}\ \bibnamefont
  {Andrade}}, \ and\ \bibinfo {author} {\bibfnamefont {Matthias}\ \bibnamefont
  {Vojta}},\ }\bibfield  {title} {\enquote {\bibinfo {title} {{Magnetization
  processes of zigzag states on the honeycomb lattice: Identifying spin models
  for $\ensuremath{\alpha}\text{\ensuremath{-}}{\mathrm{RuCl}}_{3}$ and
  ${\mathrm{Na}}_{2}{\mathrm{IrO}}_{3}$}},}\ }\href {\doibase
  10.1103/PhysRevB.96.064430} {\bibfield  {journal} {\bibinfo  {journal} {Phys.
  Rev. B}\ }\textbf {\bibinfo {volume} {96}},\ \bibinfo {pages} {064430}
  (\bibinfo {year} {2017})}\BibitemShut {NoStop}%
\bibitem [{\citenamefont {Kim}\ \emph {et~al.}(2020)\citenamefont {Kim},
  \citenamefont {Sota}, \citenamefont {Shirakawa}, \citenamefont {Yunoki},\
  and\ \citenamefont {Son}}]{PhysRevB.102.140402}%
  \BibitemOpen
  \bibfield  {author} {\bibinfo {author} {\bibfnamefont {Beom~Hyun}\
  \bibnamefont {Kim}}, \bibinfo {author} {\bibfnamefont {Shigetoshi}\
  \bibnamefont {Sota}}, \bibinfo {author} {\bibfnamefont {Tomonori}\
  \bibnamefont {Shirakawa}}, \bibinfo {author} {\bibfnamefont {Seiji}\
  \bibnamefont {Yunoki}}, \ and\ \bibinfo {author} {\bibfnamefont {Young-Woo}\
  \bibnamefont {Son}},\ }\bibfield  {title} {\enquote {\bibinfo {title}
  {{Proximate Kitaev system for an intermediate magnetic phase in in-plane
  magnetic fields}},}\ }\href {\doibase 10.1103/PhysRevB.102.140402} {\bibfield
   {journal} {\bibinfo  {journal} {Phys. Rev. B}\ }\textbf {\bibinfo {volume}
  {102}},\ \bibinfo {pages} {140402} (\bibinfo {year} {2020})}\BibitemShut
  {NoStop}%
\bibitem [{\citenamefont {Gordon}\ \emph {et~al.}(2019)\citenamefont {Gordon},
  \citenamefont {Catuneanu}, \citenamefont {S{\o}rensen},\ and\ \citenamefont
  {Kee}}]{nc10_2470}%
  \BibitemOpen
  \bibfield  {author} {\bibinfo {author} {\bibfnamefont {Jacob~S.}\
  \bibnamefont {Gordon}}, \bibinfo {author} {\bibfnamefont {Andrei}\
  \bibnamefont {Catuneanu}}, \bibinfo {author} {\bibfnamefont {Erik~S.}\
  \bibnamefont {S{\o}rensen}}, \ and\ \bibinfo {author} {\bibfnamefont
  {Hae-Young}\ \bibnamefont {Kee}},\ }\bibfield  {title} {\enquote {\bibinfo
  {title} {{Theory of the field-revealed Kitaev spin liquid}},}\ }\href@noop {}
  {\bibfield  {journal} {\bibinfo  {journal} {Nat. Commun.}\ }\textbf {\bibinfo
  {volume} {10}},\ \bibinfo {pages} {2470} (\bibinfo {year}
  {2019})}\BibitemShut {NoStop}%
\bibitem [{\citenamefont {Winter}\ \emph {et~al.}(2018)\citenamefont {Winter},
  \citenamefont {Riedl}, \citenamefont {Kaib}, \citenamefont {Coldea},\ and\
  \citenamefont {Valent\'{\i}}}]{PhysRevLett.120.077203}%
  \BibitemOpen
  \bibfield  {author} {\bibinfo {author} {\bibfnamefont {Stephen~M.}\
  \bibnamefont {Winter}}, \bibinfo {author} {\bibfnamefont {Kira}\ \bibnamefont
  {Riedl}}, \bibinfo {author} {\bibfnamefont {David}\ \bibnamefont {Kaib}},
  \bibinfo {author} {\bibfnamefont {Radu}\ \bibnamefont {Coldea}}, \ and\
  \bibinfo {author} {\bibfnamefont {Roser}\ \bibnamefont {Valent\'{\i}}},\
  }\bibfield  {title} {\enquote {\bibinfo {title} {{Probing
  $\ensuremath{\alpha}\ensuremath{-}{\mathrm{RuCl}}_{3}$ Beyond Magnetic Order:
  Effects of Temperature and Magnetic Field}},}\ }\href {\doibase
  10.1103/PhysRevLett.120.077203} {\bibfield  {journal} {\bibinfo  {journal}
  {Phys. Rev. Lett.}\ }\textbf {\bibinfo {volume} {120}},\ \bibinfo {pages}
  {077203} (\bibinfo {year} {2018})}\BibitemShut {NoStop}%
\bibitem [{\citenamefont {Janssen}\ and\ \citenamefont
  {Vojta}(2019)}]{Janssen_2019}%
  \BibitemOpen
  \bibfield  {author} {\bibinfo {author} {\bibfnamefont {Lukas}\ \bibnamefont
  {Janssen}}\ and\ \bibinfo {author} {\bibfnamefont {Matthias}\ \bibnamefont
  {Vojta}},\ }\bibfield  {title} {\enquote {\bibinfo {title}
  {{Heisenberg{\textendash}Kitaev physics in magnetic fields}},}\ }\href
  {\doibase 10.1088/1361-648x/ab283e} {\bibfield  {journal} {\bibinfo
  {journal} {J. Phys. Condens. Matter}\ }\textbf {\bibinfo {volume} {31}},\
  \bibinfo {pages} {423002} (\bibinfo {year} {2019})}\BibitemShut {NoStop}%
\bibitem [{\citenamefont {Chern}\ \emph {et~al.}(2020)\citenamefont {Chern},
  \citenamefont {Kaneko}, \citenamefont {Lee},\ and\ \citenamefont
  {Kim}}]{PhysRevResearch.2.013014}%
  \BibitemOpen
  \bibfield  {author} {\bibinfo {author} {\bibfnamefont {Li~Ern}\ \bibnamefont
  {Chern}}, \bibinfo {author} {\bibfnamefont {Ryui}\ \bibnamefont {Kaneko}},
  \bibinfo {author} {\bibfnamefont {Hyun-Yong}\ \bibnamefont {Lee}}, \ and\
  \bibinfo {author} {\bibfnamefont {Yong~Baek}\ \bibnamefont {Kim}},\
  }\bibfield  {title} {\enquote {\bibinfo {title} {{Magnetic field induced
  competing phases in spin-orbital entangled Kitaev magnets}},}\ }\href
  {\doibase 10.1103/PhysRevResearch.2.013014} {\bibfield  {journal} {\bibinfo
  {journal} {Phys. Rev. Research}\ }\textbf {\bibinfo {volume} {2}},\ \bibinfo
  {pages} {013014} (\bibinfo {year} {2020})}\BibitemShut {NoStop}%
\bibitem [{\citenamefont {Suzuki}\ \emph {et~al.}(2021)\citenamefont {Suzuki},
  \citenamefont {Liu}, \citenamefont {Bertinshaw}, \citenamefont {Ueda},
  \citenamefont {Kim}, \citenamefont {Laha}, \citenamefont {Weber},
  \citenamefont {Yang}, \citenamefont {Wang}, \citenamefont {Takahashi},
  \citenamefont {F{\"u}rsich}, \citenamefont {Minola}, \citenamefont {Lotsch},
  \citenamefont {Kim}, \citenamefont {Yava{\c s}}, \citenamefont {Daghofer},
  \citenamefont {Chaloupka}, \citenamefont {Khaliullin}, \citenamefont
  {Gretarsson},\ and\ \citenamefont {Keimer}}]{nc12_4512}%
  \BibitemOpen
  \bibfield  {author} {\bibinfo {author} {\bibfnamefont {H.}~\bibnamefont
  {Suzuki}}, \bibinfo {author} {\bibfnamefont {H.}~\bibnamefont {Liu}},
  \bibinfo {author} {\bibfnamefont {J.}~\bibnamefont {Bertinshaw}}, \bibinfo
  {author} {\bibfnamefont {K.}~\bibnamefont {Ueda}}, \bibinfo {author}
  {\bibfnamefont {H.}~\bibnamefont {Kim}}, \bibinfo {author} {\bibfnamefont
  {S.}~\bibnamefont {Laha}}, \bibinfo {author} {\bibfnamefont {D.}~\bibnamefont
  {Weber}}, \bibinfo {author} {\bibfnamefont {Z.}~\bibnamefont {Yang}},
  \bibinfo {author} {\bibfnamefont {L.}~\bibnamefont {Wang}}, \bibinfo {author}
  {\bibfnamefont {H.}~\bibnamefont {Takahashi}}, \bibinfo {author}
  {\bibfnamefont {K.}~\bibnamefont {F{\"u}rsich}}, \bibinfo {author}
  {\bibfnamefont {M.}~\bibnamefont {Minola}}, \bibinfo {author} {\bibfnamefont
  {B.~V.}\ \bibnamefont {Lotsch}}, \bibinfo {author} {\bibfnamefont {B.~J.}\
  \bibnamefont {Kim}}, \bibinfo {author} {\bibfnamefont {H.}~\bibnamefont
  {Yava{\c s}}}, \bibinfo {author} {\bibfnamefont {M.}~\bibnamefont
  {Daghofer}}, \bibinfo {author} {\bibfnamefont {J.}~\bibnamefont {Chaloupka}},
  \bibinfo {author} {\bibfnamefont {G.}~\bibnamefont {Khaliullin}}, \bibinfo
  {author} {\bibfnamefont {H.}~\bibnamefont {Gretarsson}}, \ and\ \bibinfo
  {author} {\bibfnamefont {B.}~\bibnamefont {Keimer}},\ }\bibfield  {title}
  {\enquote {\bibinfo {title} {{Proximate ferromagnetic state in the Kitaev
  model material
  $\ensuremath{\alpha}\text{\ensuremath{-}}{\mathrm{RuCl}}_{3}$}},}\
  }\href@noop {} {\bibfield  {journal} {\bibinfo  {journal} {Nat. Commun.}\
  }\textbf {\bibinfo {volume} {12}},\ \bibinfo {pages} {4512} (\bibinfo {year}
  {2021})}\BibitemShut {NoStop}%
\bibitem [{\citenamefont {Li}\ \emph {et~al.}(2021)\citenamefont {Li},
  \citenamefont {Zhang}, \citenamefont {Wang}, \citenamefont {Wu},
  \citenamefont {Gao}, \citenamefont {Qu}, \citenamefont {Liu}, \citenamefont
  {Gong},\ and\ \citenamefont {Li}}]{nc12_4007}%
  \BibitemOpen
  \bibfield  {author} {\bibinfo {author} {\bibfnamefont {Han}\ \bibnamefont
  {Li}}, \bibinfo {author} {\bibfnamefont {Hao-Kai}\ \bibnamefont {Zhang}},
  \bibinfo {author} {\bibfnamefont {Jiucai}\ \bibnamefont {Wang}}, \bibinfo
  {author} {\bibfnamefont {Han-Qing}\ \bibnamefont {Wu}}, \bibinfo {author}
  {\bibfnamefont {Yuan}\ \bibnamefont {Gao}}, \bibinfo {author} {\bibfnamefont
  {Dai-Wei}\ \bibnamefont {Qu}}, \bibinfo {author} {\bibfnamefont {Zheng-Xin}\
  \bibnamefont {Liu}}, \bibinfo {author} {\bibfnamefont {Shou-Shu}\
  \bibnamefont {Gong}}, \ and\ \bibinfo {author} {\bibfnamefont {Wei}\
  \bibnamefont {Li}},\ }\bibfield  {title} {\enquote {\bibinfo {title}
  {{Identification of magnetic interactions and high-field quantum spin liquid
  in $\ensuremath{\alpha}\text{\ensuremath{-}}{\mathrm{RuCl}}_{3}$}},}\
  }\href@noop {} {\bibfield  {journal} {\bibinfo  {journal} {Nat. Commun.}\
  }\textbf {\bibinfo {volume} {12}},\ \bibinfo {pages} {4007} (\bibinfo {year}
  {2021})}\BibitemShut {NoStop}%
\bibitem [{\citenamefont {Kubota}\ \emph {et~al.}(2015)\citenamefont {Kubota},
  \citenamefont {Tanaka}, \citenamefont {Ono}, \citenamefont {Narumi},\ and\
  \citenamefont {Kindo}}]{PhysRevB.91.094422}%
  \BibitemOpen
  \bibfield  {author} {\bibinfo {author} {\bibfnamefont {Yumi}\ \bibnamefont
  {Kubota}}, \bibinfo {author} {\bibfnamefont {Hidekazu}\ \bibnamefont
  {Tanaka}}, \bibinfo {author} {\bibfnamefont {Toshio}\ \bibnamefont {Ono}},
  \bibinfo {author} {\bibfnamefont {Yasuo}\ \bibnamefont {Narumi}}, \ and\
  \bibinfo {author} {\bibfnamefont {Koichi}\ \bibnamefont {Kindo}},\ }\bibfield
   {title} {\enquote {\bibinfo {title} {{Successive magnetic phase transitions
  in $\ensuremath{\alpha}-{\mathrm{RuCl}}_{3}$: XY-like frustrated magnet on
  the honeycomb lattice}},}\ }\href {\doibase 10.1103/PhysRevB.91.094422}
  {\bibfield  {journal} {\bibinfo  {journal} {Phys. Rev. B}\ }\textbf {\bibinfo
  {volume} {91}},\ \bibinfo {pages} {094422} (\bibinfo {year}
  {2015})}\BibitemShut {NoStop}%
\bibitem [{\citenamefont {Zheng}\ \emph {et~al.}(2017)\citenamefont {Zheng},
  \citenamefont {Ran}, \citenamefont {Li}, \citenamefont {Wang}, \citenamefont
  {Wang}, \citenamefont {Liu}, \citenamefont {Liu}, \citenamefont {Normand},
  \citenamefont {Wen},\ and\ \citenamefont {Yu}}]{PhysRevLett.119.227208}%
  \BibitemOpen
  \bibfield  {author} {\bibinfo {author} {\bibfnamefont {Jiacheng}\
  \bibnamefont {Zheng}}, \bibinfo {author} {\bibfnamefont {Kejing}\
  \bibnamefont {Ran}}, \bibinfo {author} {\bibfnamefont {Tianrun}\ \bibnamefont
  {Li}}, \bibinfo {author} {\bibfnamefont {Jinghui}\ \bibnamefont {Wang}},
  \bibinfo {author} {\bibfnamefont {Pengshuai}\ \bibnamefont {Wang}}, \bibinfo
  {author} {\bibfnamefont {Bin}\ \bibnamefont {Liu}}, \bibinfo {author}
  {\bibfnamefont {Zheng-Xin}\ \bibnamefont {Liu}}, \bibinfo {author}
  {\bibfnamefont {B.}~\bibnamefont {Normand}}, \bibinfo {author} {\bibfnamefont
  {Jinsheng}\ \bibnamefont {Wen}}, \ and\ \bibinfo {author} {\bibfnamefont
  {Weiqiang}\ \bibnamefont {Yu}},\ }\bibfield  {title} {\enquote {\bibinfo
  {title} {{Gapless Spin Excitations in the Field-Induced Quantum Spin Liquid
  Phase of $\ensuremath{\alpha}\text{\ensuremath{-}}{\mathrm{RuCl}}_{3}$}},}\
  }\href {\doibase 10.1103/PhysRevLett.119.227208} {\bibfield  {journal}
  {\bibinfo  {journal} {Phys. Rev. Lett.}\ }\textbf {\bibinfo {volume} {119}},\
  \bibinfo {pages} {227208} (\bibinfo {year} {2017})}\BibitemShut {NoStop}%
\bibitem [{\citenamefont {Majumder}\ \emph {et~al.}(2015)\citenamefont
  {Majumder}, \citenamefont {Schmidt}, \citenamefont {Rosner}, \citenamefont
  {Tsirlin}, \citenamefont {Yasuoka},\ and\ \citenamefont
  {Baenitz}}]{PhysRevB.91.180401}%
  \BibitemOpen
  \bibfield  {author} {\bibinfo {author} {\bibfnamefont {M.}~\bibnamefont
  {Majumder}}, \bibinfo {author} {\bibfnamefont {M.}~\bibnamefont {Schmidt}},
  \bibinfo {author} {\bibfnamefont {H.}~\bibnamefont {Rosner}}, \bibinfo
  {author} {\bibfnamefont {A.~A.}\ \bibnamefont {Tsirlin}}, \bibinfo {author}
  {\bibfnamefont {H.}~\bibnamefont {Yasuoka}}, \ and\ \bibinfo {author}
  {\bibfnamefont {M.}~\bibnamefont {Baenitz}},\ }\bibfield  {title} {\enquote
  {\bibinfo {title} {{Anisotropic ${\mathrm{Ru}}^{3+} 4{d}^{5}$ magnetism in
  the $\ensuremath{\alpha}-{\mathrm{RuCl}}_{3}$ honeycomb system:
  Susceptibility, specific heat, and zero-field NMR}},}\ }\href {\doibase
  10.1103/PhysRevB.91.180401} {\bibfield  {journal} {\bibinfo  {journal} {Phys.
  Rev. B}\ }\textbf {\bibinfo {volume} {91}},\ \bibinfo {pages} {180401}
  (\bibinfo {year} {2015})}\BibitemShut {NoStop}%
\bibitem [{\citenamefont {Balz}\ \emph {et~al.}(2021)\citenamefont {Balz},
  \citenamefont {Janssen}, \citenamefont {Lampen-Kelley}, \citenamefont
  {Banerjee}, \citenamefont {Liu}, \citenamefont {Yan}, \citenamefont
  {Mandrus}, \citenamefont {Vojta},\ and\ \citenamefont
  {Nagler}}]{PhysRevB.103.174417}%
  \BibitemOpen
  \bibfield  {author} {\bibinfo {author} {\bibfnamefont {C.}~\bibnamefont
  {Balz}}, \bibinfo {author} {\bibfnamefont {L.}~\bibnamefont {Janssen}},
  \bibinfo {author} {\bibfnamefont {P.}~\bibnamefont {Lampen-Kelley}}, \bibinfo
  {author} {\bibfnamefont {A.}~\bibnamefont {Banerjee}}, \bibinfo {author}
  {\bibfnamefont {Y.~H.}\ \bibnamefont {Liu}}, \bibinfo {author} {\bibfnamefont
  {J.-Q.}\ \bibnamefont {Yan}}, \bibinfo {author} {\bibfnamefont {D.~G.}\
  \bibnamefont {Mandrus}}, \bibinfo {author} {\bibfnamefont {M.}~\bibnamefont
  {Vojta}}, \ and\ \bibinfo {author} {\bibfnamefont {S.~E.}\ \bibnamefont
  {Nagler}},\ }\bibfield  {title} {\enquote {\bibinfo {title} {{Field-induced
  intermediate ordered phase and anisotropic interlayer interactions in
  $\ensuremath{\alpha}\text{\ensuremath{-}}{\mathrm{RuCl}}_{3}$}},}\ }\href
  {\doibase 10.1103/PhysRevB.103.174417} {\bibfield  {journal} {\bibinfo
  {journal} {Phys. Rev. B}\ }\textbf {\bibinfo {volume} {103}},\ \bibinfo
  {pages} {174417} (\bibinfo {year} {2021})}\BibitemShut {NoStop}%
\bibitem [{\citenamefont {Sears}\ \emph {et~al.}(2017)\citenamefont {Sears},
  \citenamefont {Zhao}, \citenamefont {Xu}, \citenamefont {Lynn},\ and\
  \citenamefont {Kim}}]{PhysRevB.95.180411}%
  \BibitemOpen
  \bibfield  {author} {\bibinfo {author} {\bibfnamefont {J.~A.}\ \bibnamefont
  {Sears}}, \bibinfo {author} {\bibfnamefont {Y.}~\bibnamefont {Zhao}},
  \bibinfo {author} {\bibfnamefont {Z.}~\bibnamefont {Xu}}, \bibinfo {author}
  {\bibfnamefont {J.~W.}\ \bibnamefont {Lynn}}, \ and\ \bibinfo {author}
  {\bibfnamefont {Young-June}\ \bibnamefont {Kim}},\ }\bibfield  {title}
  {\enquote {\bibinfo {title} {{Phase diagram of
  $\ensuremath{\alpha}\ensuremath{-}{\mathrm{RuCl}}_{3}$ in an in-plane
  magnetic field}},}\ }\href {\doibase 10.1103/PhysRevB.95.180411} {\bibfield
  {journal} {\bibinfo  {journal} {Phys. Rev. B}\ }\textbf {\bibinfo {volume}
  {95}},\ \bibinfo {pages} {180411} (\bibinfo {year} {2017})}\BibitemShut
  {NoStop}%
\bibitem [{\citenamefont {Bachus}\ \emph {et~al.}(2020)\citenamefont {Bachus},
  \citenamefont {Kaib}, \citenamefont {Tokiwa}, \citenamefont {Jesche},
  \citenamefont {Tsurkan}, \citenamefont {Loidl}, \citenamefont {Winter},
  \citenamefont {Tsirlin}, \citenamefont {Valent\'{\i}},\ and\ \citenamefont
  {Gegenwart}}]{PhysRevLett.125.097203}%
  \BibitemOpen
  \bibfield  {author} {\bibinfo {author} {\bibfnamefont {S.}~\bibnamefont
  {Bachus}}, \bibinfo {author} {\bibfnamefont {D.~A.~S.}\ \bibnamefont {Kaib}},
  \bibinfo {author} {\bibfnamefont {Y.}~\bibnamefont {Tokiwa}}, \bibinfo
  {author} {\bibfnamefont {A.}~\bibnamefont {Jesche}}, \bibinfo {author}
  {\bibfnamefont {V.}~\bibnamefont {Tsurkan}}, \bibinfo {author} {\bibfnamefont
  {A.}~\bibnamefont {Loidl}}, \bibinfo {author} {\bibfnamefont {S.~M.}\
  \bibnamefont {Winter}}, \bibinfo {author} {\bibfnamefont {A.~A.}\
  \bibnamefont {Tsirlin}}, \bibinfo {author} {\bibfnamefont {R.}~\bibnamefont
  {Valent\'{\i}}}, \ and\ \bibinfo {author} {\bibfnamefont {P.}~\bibnamefont
  {Gegenwart}},\ }\bibfield  {title} {\enquote {\bibinfo {title}
  {{Thermodynamic Perspective on Field-Induced Behavior of
  $\ensuremath{\alpha}\text{\ensuremath{-}}{\mathrm{RuCl}}_{3}$}},}\ }\href
  {\doibase 10.1103/PhysRevLett.125.097203} {\bibfield  {journal} {\bibinfo
  {journal} {Phys. Rev. Lett.}\ }\textbf {\bibinfo {volume} {125}},\ \bibinfo
  {pages} {097203} (\bibinfo {year} {2020})}\BibitemShut {NoStop}%
\bibitem [{\citenamefont {Wolter}\ \emph {et~al.}(2017)\citenamefont {Wolter},
  \citenamefont {Corredor}, \citenamefont {Janssen}, \citenamefont {Nenkov},
  \citenamefont {Sch\"onecker}, \citenamefont {Do}, \citenamefont {Choi},
  \citenamefont {Albrecht}, \citenamefont {Hunger}, \citenamefont {Doert},
  \citenamefont {Vojta},\ and\ \citenamefont {B\"uchner}}]{PhysRevB.96.041405}%
  \BibitemOpen
  \bibfield  {author} {\bibinfo {author} {\bibfnamefont {A.~U.~B.}\
  \bibnamefont {Wolter}}, \bibinfo {author} {\bibfnamefont {L.~T.}\
  \bibnamefont {Corredor}}, \bibinfo {author} {\bibfnamefont {L.}~\bibnamefont
  {Janssen}}, \bibinfo {author} {\bibfnamefont {K.}~\bibnamefont {Nenkov}},
  \bibinfo {author} {\bibfnamefont {S.}~\bibnamefont {Sch\"onecker}}, \bibinfo
  {author} {\bibfnamefont {S.-H.}\ \bibnamefont {Do}}, \bibinfo {author}
  {\bibfnamefont {K.-Y.}\ \bibnamefont {Choi}}, \bibinfo {author}
  {\bibfnamefont {R.}~\bibnamefont {Albrecht}}, \bibinfo {author}
  {\bibfnamefont {J.}~\bibnamefont {Hunger}}, \bibinfo {author} {\bibfnamefont
  {T.}~\bibnamefont {Doert}}, \bibinfo {author} {\bibfnamefont
  {M.}~\bibnamefont {Vojta}}, \ and\ \bibinfo {author} {\bibfnamefont
  {B.}~\bibnamefont {B\"uchner}},\ }\bibfield  {title} {\enquote {\bibinfo
  {title} {{Field-induced quantum criticality in the Kitaev system
  $\ensuremath{\alpha}\ensuremath{-}{\mathrm{RuCl}}_{3}$}},}\ }\href {\doibase
  10.1103/PhysRevB.96.041405} {\bibfield  {journal} {\bibinfo  {journal} {Phys.
  Rev. B}\ }\textbf {\bibinfo {volume} {96}},\ \bibinfo {pages} {041405}
  (\bibinfo {year} {2017})}\BibitemShut {NoStop}%
\bibitem [{\citenamefont {Baek}\ \emph {et~al.}(2017)\citenamefont {Baek},
  \citenamefont {Do}, \citenamefont {Choi}, \citenamefont {Kwon}, \citenamefont
  {Wolter}, \citenamefont {Nishimoto}, \citenamefont {van~den Brink},\ and\
  \citenamefont {B\"uchner}}]{PhysRevLett.119.037201}%
  \BibitemOpen
  \bibfield  {author} {\bibinfo {author} {\bibfnamefont {S.-H.}\ \bibnamefont
  {Baek}}, \bibinfo {author} {\bibfnamefont {S.-H.}\ \bibnamefont {Do}},
  \bibinfo {author} {\bibfnamefont {K.-Y.}\ \bibnamefont {Choi}}, \bibinfo
  {author} {\bibfnamefont {Y.~S.}\ \bibnamefont {Kwon}}, \bibinfo {author}
  {\bibfnamefont {A.~U.~B.}\ \bibnamefont {Wolter}}, \bibinfo {author}
  {\bibfnamefont {S.}~\bibnamefont {Nishimoto}}, \bibinfo {author}
  {\bibfnamefont {Jeroen}\ \bibnamefont {van~den Brink}}, \ and\ \bibinfo
  {author} {\bibfnamefont {B.}~\bibnamefont {B\"uchner}},\ }\bibfield  {title}
  {\enquote {\bibinfo {title} {{Evidence for a Field-Induced Quantum Spin
  Liquid in $\ensuremath{\alpha}\text{\ensuremath{-}}{\mathrm{RuCl}}_{3}$}},}\
  }\href {\doibase 10.1103/PhysRevLett.119.037201} {\bibfield  {journal}
  {\bibinfo  {journal} {Phys. Rev. Lett.}\ }\textbf {\bibinfo {volume} {119}},\
  \bibinfo {pages} {037201} (\bibinfo {year} {2017})}\BibitemShut {NoStop}%
\bibitem [{\citenamefont {Yu}\ \emph {et~al.}(2018)\citenamefont {Yu},
  \citenamefont {Xu}, \citenamefont {Ran}, \citenamefont {Ni}, \citenamefont
  {Huang}, \citenamefont {Wang}, \citenamefont {Wen},\ and\ \citenamefont
  {Li}}]{PhysRevLett.120.067202}%
  \BibitemOpen
  \bibfield  {author} {\bibinfo {author} {\bibfnamefont {Y.~J.}\ \bibnamefont
  {Yu}}, \bibinfo {author} {\bibfnamefont {Y.}~\bibnamefont {Xu}}, \bibinfo
  {author} {\bibfnamefont {K.~J.}\ \bibnamefont {Ran}}, \bibinfo {author}
  {\bibfnamefont {J.~M.}\ \bibnamefont {Ni}}, \bibinfo {author} {\bibfnamefont
  {Y.~Y.}\ \bibnamefont {Huang}}, \bibinfo {author} {\bibfnamefont {J.~H.}\
  \bibnamefont {Wang}}, \bibinfo {author} {\bibfnamefont {J.~S.}\ \bibnamefont
  {Wen}}, \ and\ \bibinfo {author} {\bibfnamefont {S.~Y.}\ \bibnamefont {Li}},\
  }\bibfield  {title} {\enquote {\bibinfo {title} {{Ultralow-Temperature
  Thermal Conductivity of the Kitaev Honeycomb Magnet
  $\ensuremath{\alpha}\text{\ensuremath{-}}{\mathrm{RuCl}}_{3}$ across the
  Field-Induced Phase Transition}},}\ }\href {\doibase
  10.1103/PhysRevLett.120.067202} {\bibfield  {journal} {\bibinfo  {journal}
  {Phys. Rev. Lett.}\ }\textbf {\bibinfo {volume} {120}},\ \bibinfo {pages}
  {067202} (\bibinfo {year} {2018})}\BibitemShut {NoStop}%
\bibitem [{\citenamefont {Bachus}\ \emph {et~al.}(2021)\citenamefont {Bachus},
  \citenamefont {Kaib}, \citenamefont {Jesche}, \citenamefont {Tsurkan},
  \citenamefont {Loidl}, \citenamefont {Winter}, \citenamefont {Tsirlin},
  \citenamefont {Valent\'{\i}},\ and\ \citenamefont
  {Gegenwart}}]{PhysRevB.103.054440}%
  \BibitemOpen
  \bibfield  {author} {\bibinfo {author} {\bibfnamefont {S.}~\bibnamefont
  {Bachus}}, \bibinfo {author} {\bibfnamefont {D.~A.~S.}\ \bibnamefont {Kaib}},
  \bibinfo {author} {\bibfnamefont {A.}~\bibnamefont {Jesche}}, \bibinfo
  {author} {\bibfnamefont {V.}~\bibnamefont {Tsurkan}}, \bibinfo {author}
  {\bibfnamefont {A.}~\bibnamefont {Loidl}}, \bibinfo {author} {\bibfnamefont
  {S.~M.}\ \bibnamefont {Winter}}, \bibinfo {author} {\bibfnamefont {A.~A.}\
  \bibnamefont {Tsirlin}}, \bibinfo {author} {\bibfnamefont {R.}~\bibnamefont
  {Valent\'{\i}}}, \ and\ \bibinfo {author} {\bibfnamefont {P.}~\bibnamefont
  {Gegenwart}},\ }\bibfield  {title} {\enquote {\bibinfo {title}
  {{Angle-dependent thermodynamics of
  $\ensuremath{\alpha}\text{\ensuremath{-}}\mathrm{Ru}{\mathrm{Cl}}_{3}$}},}\
  }\href {\doibase 10.1103/PhysRevB.103.054440} {\bibfield  {journal} {\bibinfo
   {journal} {Phys. Rev. B}\ }\textbf {\bibinfo {volume} {103}},\ \bibinfo
  {pages} {054440} (\bibinfo {year} {2021})}\BibitemShut {NoStop}%
\bibitem [{\citenamefont {Widmann}\ \emph {et~al.}(2019)\citenamefont
  {Widmann}, \citenamefont {Tsurkan}, \citenamefont {Prishchenko},
  \citenamefont {Mazurenko}, \citenamefont {Tsirlin},\ and\ \citenamefont
  {Loidl}}]{PhysRevB.99.094415}%
  \BibitemOpen
  \bibfield  {author} {\bibinfo {author} {\bibfnamefont {S.}~\bibnamefont
  {Widmann}}, \bibinfo {author} {\bibfnamefont {V.}~\bibnamefont {Tsurkan}},
  \bibinfo {author} {\bibfnamefont {D.~A.}\ \bibnamefont {Prishchenko}},
  \bibinfo {author} {\bibfnamefont {V.~G.}\ \bibnamefont {Mazurenko}}, \bibinfo
  {author} {\bibfnamefont {A.~A.}\ \bibnamefont {Tsirlin}}, \ and\ \bibinfo
  {author} {\bibfnamefont {A.}~\bibnamefont {Loidl}},\ }\bibfield  {title}
  {\enquote {\bibinfo {title} {{Thermodynamic evidence of fractionalized
  excitations in
  $\ensuremath{\alpha}\text{\ensuremath{-}}{\mathrm{RuCl}}_{3}$}},}\ }\href
  {\doibase 10.1103/PhysRevB.99.094415} {\bibfield  {journal} {\bibinfo
  {journal} {Phys. Rev. B}\ }\textbf {\bibinfo {volume} {99}},\ \bibinfo
  {pages} {094415} (\bibinfo {year} {2019})}\BibitemShut {NoStop}%
\bibitem [{\citenamefont {Tanaka}\ \emph {et~al.}(2022)\citenamefont {Tanaka},
  \citenamefont {Mizukami}, \citenamefont {Harasawa}, \citenamefont
  {Hashimoto}, \citenamefont {Hwang}, \citenamefont {Kurita}, \citenamefont
  {Tanaka}, \citenamefont {Fujimoto}, \citenamefont {Matsuda}, \citenamefont
  {Moon},\ and\ \citenamefont {Shibauchi}}]{tanaka2022thermodynamic}%
  \BibitemOpen
  \bibfield  {author} {\bibinfo {author} {\bibfnamefont {O}~\bibnamefont
  {Tanaka}}, \bibinfo {author} {\bibfnamefont {Y}~\bibnamefont {Mizukami}},
  \bibinfo {author} {\bibfnamefont {R}~\bibnamefont {Harasawa}}, \bibinfo
  {author} {\bibfnamefont {K}~\bibnamefont {Hashimoto}}, \bibinfo {author}
  {\bibfnamefont {K}~\bibnamefont {Hwang}}, \bibinfo {author} {\bibfnamefont
  {N}~\bibnamefont {Kurita}}, \bibinfo {author} {\bibfnamefont {H}~\bibnamefont
  {Tanaka}}, \bibinfo {author} {\bibfnamefont {S}~\bibnamefont {Fujimoto}},
  \bibinfo {author} {\bibfnamefont {Y}~\bibnamefont {Matsuda}}, \bibinfo
  {author} {\bibfnamefont {E-G}\ \bibnamefont {Moon}}, \ and\ \bibinfo {author}
  {\bibfnamefont {T.}~\bibnamefont {Shibauchi}},\ }\bibfield  {title} {\enquote
  {\bibinfo {title} {{Thermodynamic evidence for a field-angle-dependent
  Majorana gap in a Kitaev spin liquid}},}\ }\href {\doibase
  https://doi.org/10.1038/s41567-021-01488-6} {\bibfield  {journal} {\bibinfo
  {journal} {Nat. Phys.}\ } (\bibinfo {year} {2022}),\
  https://doi.org/10.1038/s41567-021-01488-6}\BibitemShut {NoStop}%
\bibitem [{\citenamefont {Zhou}\ \emph {et~al.}(2022)\citenamefont {Zhou},
  \citenamefont {Li}, \citenamefont {Matsuda}, \citenamefont {Matsuo},
  \citenamefont {Li}, \citenamefont {Kurita}, \citenamefont {Kindo},\ and\
  \citenamefont {Tanaka}}]{zhou2022intermediate}%
  \BibitemOpen
  \bibfield  {author} {\bibinfo {author} {\bibfnamefont {Xu-Guang}\
  \bibnamefont {Zhou}}, \bibinfo {author} {\bibfnamefont {Han}\ \bibnamefont
  {Li}}, \bibinfo {author} {\bibfnamefont {Yasuhiro~H.}\ \bibnamefont
  {Matsuda}}, \bibinfo {author} {\bibfnamefont {Akira}\ \bibnamefont {Matsuo}},
  \bibinfo {author} {\bibfnamefont {Wei}\ \bibnamefont {Li}}, \bibinfo {author}
  {\bibfnamefont {Nobuyuki}\ \bibnamefont {Kurita}}, \bibinfo {author}
  {\bibfnamefont {Koichi}\ \bibnamefont {Kindo}}, \ and\ \bibinfo {author}
  {\bibfnamefont {Hidekazu}\ \bibnamefont {Tanaka}},\ }\href@noop {} {\enquote
  {\bibinfo {title} {{Intermediate Quantum Spin Liquid Phase in the Kitaev
  Material $\alpha$-RuCl$_3$ under High Magnetic Fields up to 100 T}},}\ }
  (\bibinfo {year} {2022}),\ \Eprint {http://arxiv.org/abs/2201.04597}
  {arXiv:2201.04597} \BibitemShut {NoStop}%
\bibitem [{\citenamefont {Banerjee}\ \emph {et~al.}(2018)\citenamefont
  {Banerjee}, \citenamefont {Lampen-Kelley}, \citenamefont {Knolle},
  \citenamefont {Balz}, \citenamefont {Aczel}, \citenamefont {Winn},
  \citenamefont {Liu}, \citenamefont {Pajerowski}, \citenamefont {Yan},
  \citenamefont {Bridges}, \citenamefont {Savici}, \citenamefont {Chakoumakos},
  \citenamefont {Lumsden}, \citenamefont {Tennant}, \citenamefont {Moessner},
  \citenamefont {Mandrus},\ and\ \citenamefont {Nagler}}]{npjqm3_8}%
  \BibitemOpen
  \bibfield  {author} {\bibinfo {author} {\bibfnamefont {Arnab}\ \bibnamefont
  {Banerjee}}, \bibinfo {author} {\bibfnamefont {Paula}\ \bibnamefont
  {Lampen-Kelley}}, \bibinfo {author} {\bibfnamefont {Johannes}\ \bibnamefont
  {Knolle}}, \bibinfo {author} {\bibfnamefont {Christian}\ \bibnamefont
  {Balz}}, \bibinfo {author} {\bibfnamefont {Adam~Anthony}\ \bibnamefont
  {Aczel}}, \bibinfo {author} {\bibfnamefont {Barry}\ \bibnamefont {Winn}},
  \bibinfo {author} {\bibfnamefont {Yaohua}\ \bibnamefont {Liu}}, \bibinfo
  {author} {\bibfnamefont {Daniel}\ \bibnamefont {Pajerowski}}, \bibinfo
  {author} {\bibfnamefont {Jiaqiang}\ \bibnamefont {Yan}}, \bibinfo {author}
  {\bibfnamefont {Craig~A.}\ \bibnamefont {Bridges}}, \bibinfo {author}
  {\bibfnamefont {Andrei~T.}\ \bibnamefont {Savici}}, \bibinfo {author}
  {\bibfnamefont {Bryan~C.}\ \bibnamefont {Chakoumakos}}, \bibinfo {author}
  {\bibfnamefont {Mark~D.}\ \bibnamefont {Lumsden}}, \bibinfo {author}
  {\bibfnamefont {David~Alan}\ \bibnamefont {Tennant}}, \bibinfo {author}
  {\bibfnamefont {Roderich}\ \bibnamefont {Moessner}}, \bibinfo {author}
  {\bibfnamefont {David~G.}\ \bibnamefont {Mandrus}}, \ and\ \bibinfo {author}
  {\bibfnamefont {Stephen~E.}\ \bibnamefont {Nagler}},\ }\bibfield  {title}
  {\enquote {\bibinfo {title} {{Excitations in the field-induced quantum spin
  liquid state of
  $\ensuremath{\alpha}\text{\ensuremath{-}}{\mathrm{RuCl}}_{3}$}},}\
  }\href@noop {} {\bibfield  {journal} {\bibinfo  {journal} {npj Quant.
  Mater.}\ }\textbf {\bibinfo {volume} {3}},\ \bibinfo {pages} {8} (\bibinfo
  {year} {2018})}\BibitemShut {NoStop}%
\bibitem [{\citenamefont {Balz}\ \emph {et~al.}(2019)\citenamefont {Balz},
  \citenamefont {Lampen-Kelley}, \citenamefont {Banerjee}, \citenamefont {Yan},
  \citenamefont {Lu}, \citenamefont {Hu}, \citenamefont {Yadav}, \citenamefont
  {Takano}, \citenamefont {Liu}, \citenamefont {Tennant}, \citenamefont
  {Lumsden}, \citenamefont {Mandrus},\ and\ \citenamefont
  {Nagler}}]{PhysRevB.100.060405}%
  \BibitemOpen
  \bibfield  {author} {\bibinfo {author} {\bibfnamefont {Christian}\
  \bibnamefont {Balz}}, \bibinfo {author} {\bibfnamefont {Paula}\ \bibnamefont
  {Lampen-Kelley}}, \bibinfo {author} {\bibfnamefont {Arnab}\ \bibnamefont
  {Banerjee}}, \bibinfo {author} {\bibfnamefont {Jiaqiang}\ \bibnamefont
  {Yan}}, \bibinfo {author} {\bibfnamefont {Zhilun}\ \bibnamefont {Lu}},
  \bibinfo {author} {\bibfnamefont {Xinzhe}\ \bibnamefont {Hu}}, \bibinfo
  {author} {\bibfnamefont {Swapnil~M.}\ \bibnamefont {Yadav}}, \bibinfo
  {author} {\bibfnamefont {Yasu}\ \bibnamefont {Takano}}, \bibinfo {author}
  {\bibfnamefont {Yaohua}\ \bibnamefont {Liu}}, \bibinfo {author}
  {\bibfnamefont {D.~Alan}\ \bibnamefont {Tennant}}, \bibinfo {author}
  {\bibfnamefont {Mark~D.}\ \bibnamefont {Lumsden}}, \bibinfo {author}
  {\bibfnamefont {David}\ \bibnamefont {Mandrus}}, \ and\ \bibinfo {author}
  {\bibfnamefont {Stephen~E.}\ \bibnamefont {Nagler}},\ }\bibfield  {title}
  {\enquote {\bibinfo {title} {{Finite field regime for a quantum spin liquid
  in $\ensuremath{\alpha}\text{\ensuremath{-}}{\mathrm{RuCl}}_{3}$}},}\ }\href
  {\doibase 10.1103/PhysRevB.100.060405} {\bibfield  {journal} {\bibinfo
  {journal} {Phys. Rev. B}\ }\textbf {\bibinfo {volume} {100}},\ \bibinfo
  {pages} {060405} (\bibinfo {year} {2019})}\BibitemShut {NoStop}%
\bibitem [{\citenamefont {Nagai}\ \emph {et~al.}(2020)\citenamefont {Nagai},
  \citenamefont {Jinno}, \citenamefont {Yoshitake}, \citenamefont {Nasu},
  \citenamefont {Motome}, \citenamefont {Itoh},\ and\ \citenamefont
  {Shimizu}}]{PhysRevB.101.020414}%
  \BibitemOpen
  \bibfield  {author} {\bibinfo {author} {\bibfnamefont {Yuya}\ \bibnamefont
  {Nagai}}, \bibinfo {author} {\bibfnamefont {Takaaki}\ \bibnamefont {Jinno}},
  \bibinfo {author} {\bibfnamefont {Junki}\ \bibnamefont {Yoshitake}}, \bibinfo
  {author} {\bibfnamefont {Joji}\ \bibnamefont {Nasu}}, \bibinfo {author}
  {\bibfnamefont {Yukitoshi}\ \bibnamefont {Motome}}, \bibinfo {author}
  {\bibfnamefont {Masayuki}\ \bibnamefont {Itoh}}, \ and\ \bibinfo {author}
  {\bibfnamefont {Yasuhiro}\ \bibnamefont {Shimizu}},\ }\bibfield  {title}
  {\enquote {\bibinfo {title} {{Two-step gap opening across the quantum
  critical point in the Kitaev honeycomb magnet
  $\ensuremath{\alpha}\text{\ensuremath{-}}{\mathrm{RuCl}}_{3}$}},}\ }\href
  {\doibase 10.1103/PhysRevB.101.020414} {\bibfield  {journal} {\bibinfo
  {journal} {Phys. Rev. B}\ }\textbf {\bibinfo {volume} {101}},\ \bibinfo
  {pages} {020414} (\bibinfo {year} {2020})}\BibitemShut {NoStop}%
\bibitem [{\citenamefont {Jan{\v{s}}a}\ \emph {et~al.}(2018)\citenamefont
  {Jan{\v{s}}a}, \citenamefont {Zorko}, \citenamefont {Gomil{\v{s}}ek},
  \citenamefont {Pregelj}, \citenamefont {Kr{\"a}mer}, \citenamefont {Biner},
  \citenamefont {Biffin}, \citenamefont {R{\"u}egg},\ and\ \citenamefont
  {Klanj{\v{s}}ek}}]{np14_786}%
  \BibitemOpen
  \bibfield  {author} {\bibinfo {author} {\bibfnamefont {Nejc}\ \bibnamefont
  {Jan{\v{s}}a}}, \bibinfo {author} {\bibfnamefont {Andrej}\ \bibnamefont
  {Zorko}}, \bibinfo {author} {\bibfnamefont {Matja{\v{z}}}\ \bibnamefont
  {Gomil{\v{s}}ek}}, \bibinfo {author} {\bibfnamefont {Matej}\ \bibnamefont
  {Pregelj}}, \bibinfo {author} {\bibfnamefont {Karl~W}\ \bibnamefont
  {Kr{\"a}mer}}, \bibinfo {author} {\bibfnamefont {Daniel}\ \bibnamefont
  {Biner}}, \bibinfo {author} {\bibfnamefont {Alun}\ \bibnamefont {Biffin}},
  \bibinfo {author} {\bibfnamefont {Christian}\ \bibnamefont {R{\"u}egg}}, \
  and\ \bibinfo {author} {\bibfnamefont {Martin}\ \bibnamefont
  {Klanj{\v{s}}ek}},\ }\bibfield  {title} {\enquote {\bibinfo {title}
  {{Observation of two types of fractional excitation in the Kitaev honeycomb
  magnet}},}\ }\href@noop {} {\bibfield  {journal} {\bibinfo  {journal} {Nat.
  Phys.}\ }\textbf {\bibinfo {volume} {14}},\ \bibinfo {pages} {786--790}
  (\bibinfo {year} {2018})}\BibitemShut {NoStop}%
\bibitem [{\citenamefont {Leahy}\ \emph {et~al.}(2017)\citenamefont {Leahy},
  \citenamefont {Pocs}, \citenamefont {Siegfried}, \citenamefont {Graf},
  \citenamefont {Do}, \citenamefont {Choi}, \citenamefont {Normand},\ and\
  \citenamefont {Lee}}]{PhysRevLett.118.187203}%
  \BibitemOpen
  \bibfield  {author} {\bibinfo {author} {\bibfnamefont {Ian~A.}\ \bibnamefont
  {Leahy}}, \bibinfo {author} {\bibfnamefont {Christopher~A.}\ \bibnamefont
  {Pocs}}, \bibinfo {author} {\bibfnamefont {Peter~E.}\ \bibnamefont
  {Siegfried}}, \bibinfo {author} {\bibfnamefont {David}\ \bibnamefont {Graf}},
  \bibinfo {author} {\bibfnamefont {S.-H.}\ \bibnamefont {Do}}, \bibinfo
  {author} {\bibfnamefont {Kwang-Yong}\ \bibnamefont {Choi}}, \bibinfo {author}
  {\bibfnamefont {B.}~\bibnamefont {Normand}}, \ and\ \bibinfo {author}
  {\bibfnamefont {Minhyea}\ \bibnamefont {Lee}},\ }\bibfield  {title} {\enquote
  {\bibinfo {title} {{Anomalous Thermal Conductivity and Magnetic Torque
  Response in the Honeycomb Magnet
  $\ensuremath{\alpha}\text{\ensuremath{-}}{\mathrm{RuCl}}_{3}$}},}\ }\href
  {\doibase 10.1103/PhysRevLett.118.187203} {\bibfield  {journal} {\bibinfo
  {journal} {Phys. Rev. Lett.}\ }\textbf {\bibinfo {volume} {118}},\ \bibinfo
  {pages} {187203} (\bibinfo {year} {2017})}\BibitemShut {NoStop}%
\bibitem [{\citenamefont {Hentrich}\ \emph {et~al.}(2018)\citenamefont
  {Hentrich}, \citenamefont {Wolter}, \citenamefont {Zotos}, \citenamefont
  {Brenig}, \citenamefont {Nowak}, \citenamefont {Isaeva}, \citenamefont
  {Doert}, \citenamefont {Banerjee}, \citenamefont {Lampen-Kelley},
  \citenamefont {Mandrus}, \citenamefont {Nagler}, \citenamefont {Sears},
  \citenamefont {Kim}, \citenamefont {B\"uchner},\ and\ \citenamefont
  {Hess}}]{PhysRevLett.120.117204}%
  \BibitemOpen
  \bibfield  {author} {\bibinfo {author} {\bibfnamefont {Richard}\ \bibnamefont
  {Hentrich}}, \bibinfo {author} {\bibfnamefont {Anja U.~B.}\ \bibnamefont
  {Wolter}}, \bibinfo {author} {\bibfnamefont {Xenophon}\ \bibnamefont
  {Zotos}}, \bibinfo {author} {\bibfnamefont {Wolfram}\ \bibnamefont {Brenig}},
  \bibinfo {author} {\bibfnamefont {Domenic}\ \bibnamefont {Nowak}}, \bibinfo
  {author} {\bibfnamefont {Anna}\ \bibnamefont {Isaeva}}, \bibinfo {author}
  {\bibfnamefont {Thomas}\ \bibnamefont {Doert}}, \bibinfo {author}
  {\bibfnamefont {Arnab}\ \bibnamefont {Banerjee}}, \bibinfo {author}
  {\bibfnamefont {Paula}\ \bibnamefont {Lampen-Kelley}}, \bibinfo {author}
  {\bibfnamefont {David~G.}\ \bibnamefont {Mandrus}}, \bibinfo {author}
  {\bibfnamefont {Stephen~E.}\ \bibnamefont {Nagler}}, \bibinfo {author}
  {\bibfnamefont {Jennifer}\ \bibnamefont {Sears}}, \bibinfo {author}
  {\bibfnamefont {Young-June}\ \bibnamefont {Kim}}, \bibinfo {author}
  {\bibfnamefont {Bernd}\ \bibnamefont {B\"uchner}}, \ and\ \bibinfo {author}
  {\bibfnamefont {Christian}\ \bibnamefont {Hess}},\ }\bibfield  {title}
  {\enquote {\bibinfo {title} {{Unusual Phonon Heat Transport in
  $\ensuremath{\alpha}\text{\ensuremath{-}}{\mathrm{RuCl}}_{3}$: Strong
  Spin-Phonon Scattering and Field-Induced Spin Gap}},}\ }\href {\doibase
  10.1103/PhysRevLett.120.117204} {\bibfield  {journal} {\bibinfo  {journal}
  {Phys. Rev. Lett.}\ }\textbf {\bibinfo {volume} {120}},\ \bibinfo {pages}
  {117204} (\bibinfo {year} {2018})}\BibitemShut {NoStop}%
\bibitem [{\citenamefont {Kasahara}\ \emph
  {et~al.}(2018{\natexlab{a}})\citenamefont {Kasahara}, \citenamefont
  {Ohnishi}, \citenamefont {Mizukami}, \citenamefont {Tanaka}, \citenamefont
  {Ma}, \citenamefont {Sugii}, \citenamefont {Kurita}, \citenamefont {Tanaka},
  \citenamefont {Nasu}, \citenamefont {Motome}, \citenamefont {Shibauchi},\
  and\ \citenamefont {Matsuda}}]{nature559_227}%
  \BibitemOpen
  \bibfield  {author} {\bibinfo {author} {\bibfnamefont {Y.}~\bibnamefont
  {Kasahara}}, \bibinfo {author} {\bibfnamefont {T.}~\bibnamefont {Ohnishi}},
  \bibinfo {author} {\bibfnamefont {Y.}~\bibnamefont {Mizukami}}, \bibinfo
  {author} {\bibfnamefont {O.}~\bibnamefont {Tanaka}}, \bibinfo {author}
  {\bibfnamefont {Sixiao}\ \bibnamefont {Ma}}, \bibinfo {author} {\bibfnamefont
  {K.}~\bibnamefont {Sugii}}, \bibinfo {author} {\bibfnamefont
  {N.}~\bibnamefont {Kurita}}, \bibinfo {author} {\bibfnamefont
  {H.}~\bibnamefont {Tanaka}}, \bibinfo {author} {\bibfnamefont
  {J.}~\bibnamefont {Nasu}}, \bibinfo {author} {\bibfnamefont {Y.}~\bibnamefont
  {Motome}}, \bibinfo {author} {\bibfnamefont {T.}~\bibnamefont {Shibauchi}}, \
  and\ \bibinfo {author} {\bibfnamefont {Y.}~\bibnamefont {Matsuda}},\
  }\bibfield  {title} {\enquote {\bibinfo {title} {{Majorana quantization and
  half-integer thermal quantum Hall effect in a Kitaev spin liquid}},}\
  }\href@noop {} {\bibfield  {journal} {\bibinfo  {journal} {Nature}\ }\textbf
  {\bibinfo {volume} {559}},\ \bibinfo {pages} {227--231} (\bibinfo {year}
  {2018}{\natexlab{a}})}\BibitemShut {NoStop}%
\bibitem [{\citenamefont {Kasahara}\ \emph
  {et~al.}(2018{\natexlab{b}})\citenamefont {Kasahara}, \citenamefont {Sugii},
  \citenamefont {Ohnishi}, \citenamefont {Shimozawa}, \citenamefont
  {Yamashita}, \citenamefont {Kurita}, \citenamefont {Tanaka}, \citenamefont
  {Nasu}, \citenamefont {Motome}, \citenamefont {Shibauchi},\ and\
  \citenamefont {Matsuda}}]{PhysRevLett.120.217205}%
  \BibitemOpen
  \bibfield  {author} {\bibinfo {author} {\bibfnamefont {Y.}~\bibnamefont
  {Kasahara}}, \bibinfo {author} {\bibfnamefont {K.}~\bibnamefont {Sugii}},
  \bibinfo {author} {\bibfnamefont {T.}~\bibnamefont {Ohnishi}}, \bibinfo
  {author} {\bibfnamefont {M.}~\bibnamefont {Shimozawa}}, \bibinfo {author}
  {\bibfnamefont {M.}~\bibnamefont {Yamashita}}, \bibinfo {author}
  {\bibfnamefont {N.}~\bibnamefont {Kurita}}, \bibinfo {author} {\bibfnamefont
  {H.}~\bibnamefont {Tanaka}}, \bibinfo {author} {\bibfnamefont
  {J.}~\bibnamefont {Nasu}}, \bibinfo {author} {\bibfnamefont {Y.}~\bibnamefont
  {Motome}}, \bibinfo {author} {\bibfnamefont {T.}~\bibnamefont {Shibauchi}}, \
  and\ \bibinfo {author} {\bibfnamefont {Y.}~\bibnamefont {Matsuda}},\
  }\bibfield  {title} {\enquote {\bibinfo {title} {{Unusual Thermal Hall Effect
  in a Kitaev Spin Liquid Candidate
  $\ensuremath{\alpha}\text{\ensuremath{-}}{\mathrm{RuCl}}_{3}$}},}\ }\href
  {\doibase 10.1103/PhysRevLett.120.217205} {\bibfield  {journal} {\bibinfo
  {journal} {Phys. Rev. Lett.}\ }\textbf {\bibinfo {volume} {120}},\ \bibinfo
  {pages} {217205} (\bibinfo {year} {2018}{\natexlab{b}})}\BibitemShut
  {NoStop}%
\bibitem [{\citenamefont {Czajka}\ \emph {et~al.}(2021)\citenamefont {Czajka},
  \citenamefont {Gao}, \citenamefont {Hirschberger}, \citenamefont
  {Lampen-Kelley}, \citenamefont {Banerjee}, \citenamefont {Yan}, \citenamefont
  {Mandrus}, \citenamefont {Nagler},\ and\ \citenamefont {Ong}}]{np17_915}%
  \BibitemOpen
  \bibfield  {author} {\bibinfo {author} {\bibfnamefont {Peter}\ \bibnamefont
  {Czajka}}, \bibinfo {author} {\bibfnamefont {Tong}\ \bibnamefont {Gao}},
  \bibinfo {author} {\bibfnamefont {Max}\ \bibnamefont {Hirschberger}},
  \bibinfo {author} {\bibfnamefont {Paula}\ \bibnamefont {Lampen-Kelley}},
  \bibinfo {author} {\bibfnamefont {Arnab}\ \bibnamefont {Banerjee}}, \bibinfo
  {author} {\bibfnamefont {Jiaqiang}\ \bibnamefont {Yan}}, \bibinfo {author}
  {\bibfnamefont {David~G.}\ \bibnamefont {Mandrus}}, \bibinfo {author}
  {\bibfnamefont {Stephen~E.}\ \bibnamefont {Nagler}}, \ and\ \bibinfo {author}
  {\bibfnamefont {N.~P.}\ \bibnamefont {Ong}},\ }\bibfield  {title} {\enquote
  {\bibinfo {title} {{Oscillations of the thermal conductivity in the
  spin-liquid state of
  $\ensuremath{\alpha}\text{\ensuremath{-}}{\mathrm{RuCl}}_{3}$}},}\ }\href
  {\doibase 10.1038/s41567-021-01243-x} {\bibfield  {journal} {\bibinfo
  {journal} {Nat. Phys.}\ }\textbf {\bibinfo {volume} {17}},\ \bibinfo {pages}
  {915--919} (\bibinfo {year} {2021})}\BibitemShut {NoStop}%
\bibitem [{\citenamefont {Yokoi}\ \emph {et~al.}(2021)\citenamefont {Yokoi},
  \citenamefont {Ma}, \citenamefont {Kasahara}, \citenamefont {Kasahara},
  \citenamefont {Shibauchi}, \citenamefont {Kurita}, \citenamefont {Tanaka},
  \citenamefont {Nasu}, \citenamefont {Motome}, \citenamefont {Hickey},
  \citenamefont {Trebst},\ and\ \citenamefont
  {Matsuda}}]{doi:10.1126/science.aay5551}%
  \BibitemOpen
  \bibfield  {author} {\bibinfo {author} {\bibfnamefont {T.}~\bibnamefont
  {Yokoi}}, \bibinfo {author} {\bibfnamefont {S.}~\bibnamefont {Ma}}, \bibinfo
  {author} {\bibfnamefont {Y.}~\bibnamefont {Kasahara}}, \bibinfo {author}
  {\bibfnamefont {S.}~\bibnamefont {Kasahara}}, \bibinfo {author}
  {\bibfnamefont {T.}~\bibnamefont {Shibauchi}}, \bibinfo {author}
  {\bibfnamefont {N.}~\bibnamefont {Kurita}}, \bibinfo {author} {\bibfnamefont
  {H.}~\bibnamefont {Tanaka}}, \bibinfo {author} {\bibfnamefont
  {J.}~\bibnamefont {Nasu}}, \bibinfo {author} {\bibfnamefont {Y.}~\bibnamefont
  {Motome}}, \bibinfo {author} {\bibfnamefont {C.}~\bibnamefont {Hickey}},
  \bibinfo {author} {\bibfnamefont {S.}~\bibnamefont {Trebst}}, \ and\ \bibinfo
  {author} {\bibfnamefont {Y.}~\bibnamefont {Matsuda}},\ }\bibfield  {title}
  {\enquote {\bibinfo {title} {{Half-integer quantized anomalous thermal Hall
  effect in the Kitaev material candidate
  $\ensuremath{\alpha}\text{\ensuremath{-}}{\mathrm{RuCl}}_{3}$}},}\ }\href
  {\doibase 10.1126/science.aay5551} {\bibfield  {journal} {\bibinfo  {journal}
  {Science}\ }\textbf {\bibinfo {volume} {373}},\ \bibinfo {pages} {568--572}
  (\bibinfo {year} {2021})}\BibitemShut {NoStop}%
\bibitem [{\citenamefont {Bruin}\ \emph {et~al.}(2022)\citenamefont {Bruin},
  \citenamefont {Claus}, \citenamefont {Matsumoto}, \citenamefont {Kurita},
  \citenamefont {Tanaka},\ and\ \citenamefont {Takagi}}]{bruin2021robustness}%
  \BibitemOpen
  \bibfield  {author} {\bibinfo {author} {\bibfnamefont {J.~A.~N.}\
  \bibnamefont {Bruin}}, \bibinfo {author} {\bibfnamefont {R.~R.}\ \bibnamefont
  {Claus}}, \bibinfo {author} {\bibfnamefont {Y.}~\bibnamefont {Matsumoto}},
  \bibinfo {author} {\bibfnamefont {N.}~\bibnamefont {Kurita}}, \bibinfo
  {author} {\bibfnamefont {H.}~\bibnamefont {Tanaka}}, \ and\ \bibinfo {author}
  {\bibfnamefont {H.}~\bibnamefont {Takagi}},\ }\bibfield  {title} {\enquote
  {\bibinfo {title} {{Robustness of the thermal Hall effect close to
  half-quantization in $\alpha$-RuCl$_3$}},}\ }\href {\doibase
  https://doi.org/10.1038/s41567-021-01501-y} {\bibfield  {journal} {\bibinfo
  {journal} {Nat. Phys.}\ } (\bibinfo {year} {2022}),\
  https://doi.org/10.1038/s41567-021-01501-y}\BibitemShut {NoStop}%
\bibitem [{\citenamefont {Czajka}\ \emph {et~al.}(2022)\citenamefont {Czajka},
  \citenamefont {Gao}, \citenamefont {Hirschberger}, \citenamefont
  {Lampen-Kelley}, \citenamefont {Banerjee}, \citenamefont {Quirk},
  \citenamefont {Mandrus}, \citenamefont {Nagler},\ and\ \citenamefont
  {Ong}}]{czajka2022planar}%
  \BibitemOpen
  \bibfield  {author} {\bibinfo {author} {\bibfnamefont {Peter}\ \bibnamefont
  {Czajka}}, \bibinfo {author} {\bibfnamefont {Tong}\ \bibnamefont {Gao}},
  \bibinfo {author} {\bibfnamefont {Max}\ \bibnamefont {Hirschberger}},
  \bibinfo {author} {\bibfnamefont {Paula}\ \bibnamefont {Lampen-Kelley}},
  \bibinfo {author} {\bibfnamefont {Arnab}\ \bibnamefont {Banerjee}}, \bibinfo
  {author} {\bibfnamefont {Nicholas}\ \bibnamefont {Quirk}}, \bibinfo {author}
  {\bibfnamefont {David~G.}\ \bibnamefont {Mandrus}}, \bibinfo {author}
  {\bibfnamefont {Stephen~E.}\ \bibnamefont {Nagler}}, \ and\ \bibinfo {author}
  {\bibfnamefont {N.~P.}\ \bibnamefont {Ong}},\ }\href@noop {} {\enquote
  {\bibinfo {title} {{The planar thermal Hall conductivity in the Kitaev magnet
  $\ensuremath{\alpha}\text{\ensuremath{-}}{\mathrm{RuCl}}_{3}$}},}\ }
  (\bibinfo {year} {2022}),\ \Eprint {http://arxiv.org/abs/2201.07873}
  {arXiv:2201.07873} \BibitemShut {NoStop}%
\bibitem [{\citenamefont {Sahasrabudhe}\ \emph {et~al.}(2020)\citenamefont
  {Sahasrabudhe}, \citenamefont {Kaib}, \citenamefont {Reschke}, \citenamefont
  {German}, \citenamefont {Koethe}, \citenamefont {Buhot}, \citenamefont
  {Kamenskyi}, \citenamefont {Hickey}, \citenamefont {Becker}, \citenamefont
  {Tsurkan}, \citenamefont {Loidl}, \citenamefont {Do}, \citenamefont {Choi},
  \citenamefont {Gr\"uninger}, \citenamefont {Winter}, \citenamefont {Wang},
  \citenamefont {Valent\'{\i}},\ and\ \citenamefont {van
  Loosdrecht}}]{PhysRevB.101.140410}%
  \BibitemOpen
  \bibfield  {author} {\bibinfo {author} {\bibfnamefont {A.}~\bibnamefont
  {Sahasrabudhe}}, \bibinfo {author} {\bibfnamefont {D.~A.~S.}\ \bibnamefont
  {Kaib}}, \bibinfo {author} {\bibfnamefont {S.}~\bibnamefont {Reschke}},
  \bibinfo {author} {\bibfnamefont {R.}~\bibnamefont {German}}, \bibinfo
  {author} {\bibfnamefont {T.~C.}\ \bibnamefont {Koethe}}, \bibinfo {author}
  {\bibfnamefont {J.}~\bibnamefont {Buhot}}, \bibinfo {author} {\bibfnamefont
  {D.}~\bibnamefont {Kamenskyi}}, \bibinfo {author} {\bibfnamefont
  {C.}~\bibnamefont {Hickey}}, \bibinfo {author} {\bibfnamefont
  {P.}~\bibnamefont {Becker}}, \bibinfo {author} {\bibfnamefont
  {V.}~\bibnamefont {Tsurkan}}, \bibinfo {author} {\bibfnamefont
  {A.}~\bibnamefont {Loidl}}, \bibinfo {author} {\bibfnamefont {S.~H.}\
  \bibnamefont {Do}}, \bibinfo {author} {\bibfnamefont {K.~Y.}\ \bibnamefont
  {Choi}}, \bibinfo {author} {\bibfnamefont {M.}~\bibnamefont {Gr\"uninger}},
  \bibinfo {author} {\bibfnamefont {S.~M.}\ \bibnamefont {Winter}}, \bibinfo
  {author} {\bibfnamefont {Zhe}\ \bibnamefont {Wang}}, \bibinfo {author}
  {\bibfnamefont {R.}~\bibnamefont {Valent\'{\i}}}, \ and\ \bibinfo {author}
  {\bibfnamefont {P.~H.~M.}\ \bibnamefont {van Loosdrecht}},\ }\bibfield
  {title} {\enquote {\bibinfo {title} {{High-field quantum disordered state in
  $\ensuremath{\alpha}\ensuremath{-}{\mathrm{RuCl}}_{3}$: Spin flips, bound
  states, and multiparticle continuum}},}\ }\href {\doibase
  10.1103/PhysRevB.101.140410} {\bibfield  {journal} {\bibinfo  {journal}
  {Phys. Rev. B}\ }\textbf {\bibinfo {volume} {101}},\ \bibinfo {pages}
  {140410} (\bibinfo {year} {2020})}\BibitemShut {NoStop}%
\bibitem [{\citenamefont {Wellm}\ \emph {et~al.}(2018)\citenamefont {Wellm},
  \citenamefont {Zeisner}, \citenamefont {Alfonsov}, \citenamefont {Wolter},
  \citenamefont {Roslova}, \citenamefont {Isaeva}, \citenamefont {Doert},
  \citenamefont {Vojta}, \citenamefont {B\"uchner},\ and\ \citenamefont
  {Kataev}}]{PhysRevB.98.184408}%
  \BibitemOpen
  \bibfield  {author} {\bibinfo {author} {\bibfnamefont {C.}~\bibnamefont
  {Wellm}}, \bibinfo {author} {\bibfnamefont {J.}~\bibnamefont {Zeisner}},
  \bibinfo {author} {\bibfnamefont {A.}~\bibnamefont {Alfonsov}}, \bibinfo
  {author} {\bibfnamefont {A.~U.~B.}\ \bibnamefont {Wolter}}, \bibinfo {author}
  {\bibfnamefont {M.}~\bibnamefont {Roslova}}, \bibinfo {author} {\bibfnamefont
  {A.}~\bibnamefont {Isaeva}}, \bibinfo {author} {\bibfnamefont
  {T.}~\bibnamefont {Doert}}, \bibinfo {author} {\bibfnamefont
  {M.}~\bibnamefont {Vojta}}, \bibinfo {author} {\bibfnamefont
  {B.}~\bibnamefont {B\"uchner}}, \ and\ \bibinfo {author} {\bibfnamefont
  {V.}~\bibnamefont {Kataev}},\ }\bibfield  {title} {\enquote {\bibinfo {title}
  {{Signatures of low-energy fractionalized excitations in
  $\ensuremath{\alpha}\text{\ensuremath{-}}{\mathrm{RuCl}}_{3}$ from
  field-dependent microwave absorption}},}\ }\href {\doibase
  10.1103/PhysRevB.98.184408} {\bibfield  {journal} {\bibinfo  {journal} {Phys.
  Rev. B}\ }\textbf {\bibinfo {volume} {98}},\ \bibinfo {pages} {184408}
  (\bibinfo {year} {2018})}\BibitemShut {NoStop}%
\bibitem [{\citenamefont {Little}\ \emph {et~al.}(2017)\citenamefont {Little},
  \citenamefont {Wu}, \citenamefont {Lampen-Kelley}, \citenamefont {Banerjee},
  \citenamefont {Patankar}, \citenamefont {Rees}, \citenamefont {Bridges},
  \citenamefont {Yan}, \citenamefont {Mandrus}, \citenamefont {Nagler},\ and\
  \citenamefont {Orenstein}}]{PhysRevLett.119.227201}%
  \BibitemOpen
  \bibfield  {author} {\bibinfo {author} {\bibfnamefont {A.}~\bibnamefont
  {Little}}, \bibinfo {author} {\bibfnamefont {Liang}\ \bibnamefont {Wu}},
  \bibinfo {author} {\bibfnamefont {P.}~\bibnamefont {Lampen-Kelley}}, \bibinfo
  {author} {\bibfnamefont {A.}~\bibnamefont {Banerjee}}, \bibinfo {author}
  {\bibfnamefont {S.}~\bibnamefont {Patankar}}, \bibinfo {author}
  {\bibfnamefont {D.}~\bibnamefont {Rees}}, \bibinfo {author} {\bibfnamefont
  {C.~A.}\ \bibnamefont {Bridges}}, \bibinfo {author} {\bibfnamefont {J.-Q.}\
  \bibnamefont {Yan}}, \bibinfo {author} {\bibfnamefont {D.}~\bibnamefont
  {Mandrus}}, \bibinfo {author} {\bibfnamefont {S.~E.}\ \bibnamefont {Nagler}},
  \ and\ \bibinfo {author} {\bibfnamefont {J.}~\bibnamefont {Orenstein}},\
  }\bibfield  {title} {\enquote {\bibinfo {title} {{Antiferromagnetic Resonance
  and Terahertz Continuum in
  $\ensuremath{\alpha}\text{\ensuremath{-}}{\mathrm{RuCl}}_{3}$}},}\ }\href
  {\doibase 10.1103/PhysRevLett.119.227201} {\bibfield  {journal} {\bibinfo
  {journal} {Phys. Rev. Lett.}\ }\textbf {\bibinfo {volume} {119}},\ \bibinfo
  {pages} {227201} (\bibinfo {year} {2017})}\BibitemShut {NoStop}%
\bibitem [{\citenamefont {Wang}\ \emph
  {et~al.}(2017{\natexlab{b}})\citenamefont {Wang}, \citenamefont {Reschke},
  \citenamefont {H\"uvonen}, \citenamefont {Do}, \citenamefont {Choi},
  \citenamefont {Gensch}, \citenamefont {Nagel}, \citenamefont {R\~o\ om},\
  and\ \citenamefont {Loidl}}]{PhysRevLett.119.227202}%
  \BibitemOpen
  \bibfield  {author} {\bibinfo {author} {\bibfnamefont {Zhe}\ \bibnamefont
  {Wang}}, \bibinfo {author} {\bibfnamefont {S.}~\bibnamefont {Reschke}},
  \bibinfo {author} {\bibfnamefont {D.}~\bibnamefont {H\"uvonen}}, \bibinfo
  {author} {\bibfnamefont {S.-H.}\ \bibnamefont {Do}}, \bibinfo {author}
  {\bibfnamefont {K.-Y.}\ \bibnamefont {Choi}}, \bibinfo {author}
  {\bibfnamefont {M.}~\bibnamefont {Gensch}}, \bibinfo {author} {\bibfnamefont
  {U.}~\bibnamefont {Nagel}}, \bibinfo {author} {\bibfnamefont
  {T.}~\bibnamefont {R\~o\ om}}, \ and\ \bibinfo {author} {\bibfnamefont
  {A.}~\bibnamefont {Loidl}},\ }\bibfield  {title} {\enquote {\bibinfo {title}
  {{Magnetic Excitations and Continuum of a Possibly Field-Induced Quantum Spin
  Liquid in $\ensuremath{\alpha}\text{\ensuremath{-}}{\mathrm{RuCl}}_{3}$}},}\
  }\href {\doibase 10.1103/PhysRevLett.119.227202} {\bibfield  {journal}
  {\bibinfo  {journal} {Phys. Rev. Lett.}\ }\textbf {\bibinfo {volume} {119}},\
  \bibinfo {pages} {227202} (\bibinfo {year} {2017}{\natexlab{b}})}\BibitemShut
  {NoStop}%
\bibitem [{\citenamefont {Wu}\ \emph {et~al.}(2018)\citenamefont {Wu},
  \citenamefont {Little}, \citenamefont {Aldape}, \citenamefont {Rees},
  \citenamefont {Thewalt}, \citenamefont {Lampen-Kelley}, \citenamefont
  {Banerjee}, \citenamefont {Bridges}, \citenamefont {Yan}, \citenamefont
  {Boone}, \citenamefont {Patankar}, \citenamefont {Goldhaber-Gordon},
  \citenamefont {Mandrus}, \citenamefont {Nagler}, \citenamefont {Altman},\
  and\ \citenamefont {Orenstein}}]{PhysRevB.98.094425}%
  \BibitemOpen
  \bibfield  {author} {\bibinfo {author} {\bibfnamefont {Liang}\ \bibnamefont
  {Wu}}, \bibinfo {author} {\bibfnamefont {A.}~\bibnamefont {Little}}, \bibinfo
  {author} {\bibfnamefont {E.~E.}\ \bibnamefont {Aldape}}, \bibinfo {author}
  {\bibfnamefont {D.}~\bibnamefont {Rees}}, \bibinfo {author} {\bibfnamefont
  {E.}~\bibnamefont {Thewalt}}, \bibinfo {author} {\bibfnamefont
  {P.}~\bibnamefont {Lampen-Kelley}}, \bibinfo {author} {\bibfnamefont
  {A.}~\bibnamefont {Banerjee}}, \bibinfo {author} {\bibfnamefont {C.~A.}\
  \bibnamefont {Bridges}}, \bibinfo {author} {\bibfnamefont {J.-Q.}\
  \bibnamefont {Yan}}, \bibinfo {author} {\bibfnamefont {D.}~\bibnamefont
  {Boone}}, \bibinfo {author} {\bibfnamefont {S.}~\bibnamefont {Patankar}},
  \bibinfo {author} {\bibfnamefont {D.}~\bibnamefont {Goldhaber-Gordon}},
  \bibinfo {author} {\bibfnamefont {D.}~\bibnamefont {Mandrus}}, \bibinfo
  {author} {\bibfnamefont {S.~E.}\ \bibnamefont {Nagler}}, \bibinfo {author}
  {\bibfnamefont {E.}~\bibnamefont {Altman}}, \ and\ \bibinfo {author}
  {\bibfnamefont {J.}~\bibnamefont {Orenstein}},\ }\bibfield  {title} {\enquote
  {\bibinfo {title} {{Field evolution of magnons in
  $\ensuremath{\alpha}\text{\ensuremath{-}}{\mathrm{RuCl}}_{3}$ by
  high-resolution polarized terahertz spectroscopy}},}\ }\href {\doibase
  10.1103/PhysRevB.98.094425} {\bibfield  {journal} {\bibinfo  {journal} {Phys.
  Rev. B}\ }\textbf {\bibinfo {volume} {98}},\ \bibinfo {pages} {094425}
  (\bibinfo {year} {2018})}\BibitemShut {NoStop}%
\bibitem [{\citenamefont {Wulferding}\ \emph {et~al.}(2020)\citenamefont
  {Wulferding}, \citenamefont {Choi}, \citenamefont {Do}, \citenamefont {Lee},
  \citenamefont {Lemmens}, \citenamefont {Faugeras}, \citenamefont {Gallais},\
  and\ \citenamefont {Choi}}]{wulferding2020magnon}%
  \BibitemOpen
  \bibfield  {author} {\bibinfo {author} {\bibfnamefont {Dirk}\ \bibnamefont
  {Wulferding}}, \bibinfo {author} {\bibfnamefont {Youngsu}\ \bibnamefont
  {Choi}}, \bibinfo {author} {\bibfnamefont {Seung-Hwan}\ \bibnamefont {Do}},
  \bibinfo {author} {\bibfnamefont {Chan~Hyeon}\ \bibnamefont {Lee}}, \bibinfo
  {author} {\bibfnamefont {Peter}\ \bibnamefont {Lemmens}}, \bibinfo {author}
  {\bibfnamefont {Clement}\ \bibnamefont {Faugeras}}, \bibinfo {author}
  {\bibfnamefont {Yann}\ \bibnamefont {Gallais}}, \ and\ \bibinfo {author}
  {\bibfnamefont {Kwang-Yong}\ \bibnamefont {Choi}},\ }\bibfield  {title}
  {\enquote {\bibinfo {title} {{Magnon bound states versus anyonic Majorana
  excitations in the Kitaev honeycomb magnet
  $\ensuremath{\alpha}\text{\ensuremath{-}}{\mathrm{RuCl}}_{3}$}},}\
  }\href@noop {} {\bibfield  {journal} {\bibinfo  {journal} {Nat. Commun.}\
  }\textbf {\bibinfo {volume} {11}},\ \bibinfo {pages} {1--7} (\bibinfo {year}
  {2020})}\BibitemShut {NoStop}%
\bibitem [{\citenamefont {Aoyama}\ \emph {et~al.}(2017)\citenamefont {Aoyama},
  \citenamefont {Hasegawa}, \citenamefont {Kimura}, \citenamefont {Kimura},\
  and\ \citenamefont {Ohgushi}}]{PhysRevB.95.245104}%
  \BibitemOpen
  \bibfield  {author} {\bibinfo {author} {\bibfnamefont {Takuya}\ \bibnamefont
  {Aoyama}}, \bibinfo {author} {\bibfnamefont {Yoshinao}\ \bibnamefont
  {Hasegawa}}, \bibinfo {author} {\bibfnamefont {Shojiro}\ \bibnamefont
  {Kimura}}, \bibinfo {author} {\bibfnamefont {Tsuyoshi}\ \bibnamefont
  {Kimura}}, \ and\ \bibinfo {author} {\bibfnamefont {Kenya}\ \bibnamefont
  {Ohgushi}},\ }\bibfield  {title} {\enquote {\bibinfo {title} {{Anisotropic
  magnetodielectric effect in the honeycomb-type magnet
  $\ensuremath{\alpha}\text{\ensuremath{-}}{\mathrm{RuCl}}_{3}$}},}\ }\href
  {\doibase 10.1103/PhysRevB.95.245104} {\bibfield  {journal} {\bibinfo
  {journal} {Phys. Rev. B}\ }\textbf {\bibinfo {volume} {95}},\ \bibinfo
  {pages} {245104} (\bibinfo {year} {2017})}\BibitemShut {NoStop}%
\bibitem [{\citenamefont {Modic}\ \emph {et~al.}(2021)\citenamefont {Modic},
  \citenamefont {McDonald}, \citenamefont {Ruff}, \citenamefont {Bachmann},
  \citenamefont {Lai}, \citenamefont {Palmstrom}, \citenamefont {Graf},
  \citenamefont {Chan}, \citenamefont {Balakirev}, \citenamefont {Betts},
  \citenamefont {Boebinger}, \citenamefont {Schmidt}, \citenamefont {Lawler},
  \citenamefont {Sokolov}, \citenamefont {Moll}, \citenamefont {Ramshaw},\ and\
  \citenamefont {Shekhter}}]{modic2021scale}%
  \BibitemOpen
  \bibfield  {author} {\bibinfo {author} {\bibfnamefont {Kimberly~A}\
  \bibnamefont {Modic}}, \bibinfo {author} {\bibfnamefont {Ross~D.}\
  \bibnamefont {McDonald}}, \bibinfo {author} {\bibfnamefont {J.~P.~C}\
  \bibnamefont {Ruff}}, \bibinfo {author} {\bibfnamefont {Maja~D.}\
  \bibnamefont {Bachmann}}, \bibinfo {author} {\bibfnamefont {You}\
  \bibnamefont {Lai}}, \bibinfo {author} {\bibfnamefont {Johanna~C.}\
  \bibnamefont {Palmstrom}}, \bibinfo {author} {\bibfnamefont {David}\
  \bibnamefont {Graf}}, \bibinfo {author} {\bibfnamefont {Mun~K.}\ \bibnamefont
  {Chan}}, \bibinfo {author} {\bibfnamefont {F.~F.}\ \bibnamefont {Balakirev}},
  \bibinfo {author} {\bibfnamefont {J.~B.}\ \bibnamefont {Betts}}, \bibinfo
  {author} {\bibfnamefont {G.~S.}\ \bibnamefont {Boebinger}}, \bibinfo {author}
  {\bibfnamefont {Marcus}\ \bibnamefont {Schmidt}}, \bibinfo {author}
  {\bibfnamefont {J.~Michael}\ \bibnamefont {Lawler}}, \bibinfo {author}
  {\bibfnamefont {D.~A.}\ \bibnamefont {Sokolov}}, \bibinfo {author}
  {\bibfnamefont {Philip J.~W.}\ \bibnamefont {Moll}}, \bibinfo {author}
  {\bibfnamefont {B.~J.}\ \bibnamefont {Ramshaw}}, \ and\ \bibinfo {author}
  {\bibfnamefont {Arkady}\ \bibnamefont {Shekhter}},\ }\bibfield  {title}
  {\enquote {\bibinfo {title} {{Scale-invariant magnetic anisotropy in RuCl$_3$
  at high magnetic fields}},}\ }\href@noop {} {\bibfield  {journal} {\bibinfo
  {journal} {Nat. Phys.}\ }\textbf {\bibinfo {volume} {17}},\ \bibinfo {pages}
  {240--244} (\bibinfo {year} {2021})}\BibitemShut {NoStop}%
\bibitem [{\citenamefont {Ponomaryov}\ \emph {et~al.}(2017)\citenamefont
  {Ponomaryov}, \citenamefont {Schulze}, \citenamefont {Wosnitza},
  \citenamefont {Lampen-Kelley}, \citenamefont {Banerjee}, \citenamefont {Yan},
  \citenamefont {Bridges}, \citenamefont {Mandrus}, \citenamefont {Nagler},
  \citenamefont {Kolezhuk},\ and\ \citenamefont
  {Zvyagin}}]{PhysRevB.96.241107}%
  \BibitemOpen
  \bibfield  {author} {\bibinfo {author} {\bibfnamefont {A.~N.}\ \bibnamefont
  {Ponomaryov}}, \bibinfo {author} {\bibfnamefont {E.}~\bibnamefont {Schulze}},
  \bibinfo {author} {\bibfnamefont {J.}~\bibnamefont {Wosnitza}}, \bibinfo
  {author} {\bibfnamefont {P.}~\bibnamefont {Lampen-Kelley}}, \bibinfo {author}
  {\bibfnamefont {A.}~\bibnamefont {Banerjee}}, \bibinfo {author}
  {\bibfnamefont {J.-Q.}\ \bibnamefont {Yan}}, \bibinfo {author} {\bibfnamefont
  {C.~A.}\ \bibnamefont {Bridges}}, \bibinfo {author} {\bibfnamefont {D.~G.}\
  \bibnamefont {Mandrus}}, \bibinfo {author} {\bibfnamefont {S.~E.}\
  \bibnamefont {Nagler}}, \bibinfo {author} {\bibfnamefont {A.~K.}\
  \bibnamefont {Kolezhuk}}, \ and\ \bibinfo {author} {\bibfnamefont {S.~A.}\
  \bibnamefont {Zvyagin}},\ }\bibfield  {title} {\enquote {\bibinfo {title}
  {{Unconventional spin dynamics in the honeycomb-lattice material
  $\ensuremath{\alpha}\text{\ensuremath{-}}{\mathrm{RuCl}}_{3}$: High-field
  electron spin resonance studies}},}\ }\href {\doibase
  10.1103/PhysRevB.96.241107} {\bibfield  {journal} {\bibinfo  {journal} {Phys.
  Rev. B}\ }\textbf {\bibinfo {volume} {96}},\ \bibinfo {pages} {241107}
  (\bibinfo {year} {2017})}\BibitemShut {NoStop}%
\bibitem [{\citenamefont {Ponomaryov}\ \emph {et~al.}(2020)\citenamefont
  {Ponomaryov}, \citenamefont {Zviagina}, \citenamefont {Wosnitza},
  \citenamefont {Lampen-Kelley}, \citenamefont {Banerjee}, \citenamefont {Yan},
  \citenamefont {Bridges}, \citenamefont {Mandrus}, \citenamefont {Nagler},\
  and\ \citenamefont {Zvyagin}}]{PhysRevLett.125.037202}%
  \BibitemOpen
  \bibfield  {author} {\bibinfo {author} {\bibfnamefont {A.~N.}\ \bibnamefont
  {Ponomaryov}}, \bibinfo {author} {\bibfnamefont {L.}~\bibnamefont
  {Zviagina}}, \bibinfo {author} {\bibfnamefont {J.}~\bibnamefont {Wosnitza}},
  \bibinfo {author} {\bibfnamefont {P.}~\bibnamefont {Lampen-Kelley}}, \bibinfo
  {author} {\bibfnamefont {A.}~\bibnamefont {Banerjee}}, \bibinfo {author}
  {\bibfnamefont {J.-Q.}\ \bibnamefont {Yan}}, \bibinfo {author} {\bibfnamefont
  {C.~A.}\ \bibnamefont {Bridges}}, \bibinfo {author} {\bibfnamefont {D.~G.}\
  \bibnamefont {Mandrus}}, \bibinfo {author} {\bibfnamefont {S.~E.}\
  \bibnamefont {Nagler}}, \ and\ \bibinfo {author} {\bibfnamefont {S.~A.}\
  \bibnamefont {Zvyagin}},\ }\bibfield  {title} {\enquote {\bibinfo {title}
  {{Nature of Magnetic Excitations in the High-Field Phase of
  $\ensuremath{\alpha}\text{\ensuremath{-}}{\mathrm{RuCl}}_{3}$}},}\ }\href
  {\doibase 10.1103/PhysRevLett.125.037202} {\bibfield  {journal} {\bibinfo
  {journal} {Phys. Rev. Lett.}\ }\textbf {\bibinfo {volume} {125}},\ \bibinfo
  {pages} {037202} (\bibinfo {year} {2020})}\BibitemShut {NoStop}%
\bibitem [{\citenamefont {Gass}\ \emph {et~al.}(2020)\citenamefont {Gass},
  \citenamefont {C\^onsoli}, \citenamefont {Kocsis}, \citenamefont {Corredor},
  \citenamefont {Lampen-Kelley}, \citenamefont {Mandrus}, \citenamefont
  {Nagler}, \citenamefont {Janssen}, \citenamefont {Vojta}, \citenamefont
  {B\"uchner},\ and\ \citenamefont {Wolter}}]{PhysRevB.101.245158}%
  \BibitemOpen
  \bibfield  {author} {\bibinfo {author} {\bibfnamefont {S.}~\bibnamefont
  {Gass}}, \bibinfo {author} {\bibfnamefont {P.~M.}\ \bibnamefont {C\^onsoli}},
  \bibinfo {author} {\bibfnamefont {V.}~\bibnamefont {Kocsis}}, \bibinfo
  {author} {\bibfnamefont {L.~T.}\ \bibnamefont {Corredor}}, \bibinfo {author}
  {\bibfnamefont {P.}~\bibnamefont {Lampen-Kelley}}, \bibinfo {author}
  {\bibfnamefont {D.~G.}\ \bibnamefont {Mandrus}}, \bibinfo {author}
  {\bibfnamefont {S.~E.}\ \bibnamefont {Nagler}}, \bibinfo {author}
  {\bibfnamefont {L.}~\bibnamefont {Janssen}}, \bibinfo {author} {\bibfnamefont
  {M.}~\bibnamefont {Vojta}}, \bibinfo {author} {\bibfnamefont
  {B.}~\bibnamefont {B\"uchner}}, \ and\ \bibinfo {author} {\bibfnamefont
  {A.~U.~B.}\ \bibnamefont {Wolter}},\ }\bibfield  {title} {\enquote {\bibinfo
  {title} {{Field-induced transitions in the Kitaev material
  $\ensuremath{\alpha}\ensuremath{-}{\mathrm{RuCl}}_{3}$ probed by thermal
  expansion and magnetostriction}},}\ }\href {\doibase
  10.1103/PhysRevB.101.245158} {\bibfield  {journal} {\bibinfo  {journal}
  {Phys. Rev. B}\ }\textbf {\bibinfo {volume} {101}},\ \bibinfo {pages}
  {245158} (\bibinfo {year} {2020})}\BibitemShut {NoStop}%
\bibitem [{\citenamefont {Sch\"onemann}\ \emph {et~al.}(2020)\citenamefont
  {Sch\"onemann}, \citenamefont {Imajo}, \citenamefont {Weickert},
  \citenamefont {Yan}, \citenamefont {Mandrus}, \citenamefont {Takano},
  \citenamefont {Brosha}, \citenamefont {Rosa}, \citenamefont {Nagler},
  \citenamefont {Kindo},\ and\ \citenamefont {Jaime}}]{PhysRevB.102.214432}%
  \BibitemOpen
  \bibfield  {author} {\bibinfo {author} {\bibfnamefont {Rico}\ \bibnamefont
  {Sch\"onemann}}, \bibinfo {author} {\bibfnamefont {Shusaku}\ \bibnamefont
  {Imajo}}, \bibinfo {author} {\bibfnamefont {Franziska}\ \bibnamefont
  {Weickert}}, \bibinfo {author} {\bibfnamefont {Jiaqiang}\ \bibnamefont
  {Yan}}, \bibinfo {author} {\bibfnamefont {David~G.}\ \bibnamefont {Mandrus}},
  \bibinfo {author} {\bibfnamefont {Yasumasa}\ \bibnamefont {Takano}}, \bibinfo
  {author} {\bibfnamefont {Eric~L.}\ \bibnamefont {Brosha}}, \bibinfo {author}
  {\bibfnamefont {Priscila F.~S.}\ \bibnamefont {Rosa}}, \bibinfo {author}
  {\bibfnamefont {Stephen~E.}\ \bibnamefont {Nagler}}, \bibinfo {author}
  {\bibfnamefont {Koichi}\ \bibnamefont {Kindo}}, \ and\ \bibinfo {author}
  {\bibfnamefont {Marcelo}\ \bibnamefont {Jaime}},\ }\bibfield  {title}
  {\enquote {\bibinfo {title} {{Thermal and magnetoelastic properties of
  $\ensuremath{\alpha}\ensuremath{-}{\mathrm{RuCl}}_{3}$ in the field-induced
  low-temperature states}},}\ }\href {\doibase 10.1103/PhysRevB.102.214432}
  {\bibfield  {journal} {\bibinfo  {journal} {Phys. Rev. B}\ }\textbf {\bibinfo
  {volume} {102}},\ \bibinfo {pages} {214432} (\bibinfo {year}
  {2020})}\BibitemShut {NoStop}%
\bibitem [{\citenamefont {Maksimov}\ and\ \citenamefont
  {Chernyshev}(2020)}]{PhysRevResearch.2.033011}%
  \BibitemOpen
  \bibfield  {author} {\bibinfo {author} {\bibfnamefont {P.~A.}\ \bibnamefont
  {Maksimov}}\ and\ \bibinfo {author} {\bibfnamefont {A.~L.}\ \bibnamefont
  {Chernyshev}},\ }\bibfield  {title} {\enquote {\bibinfo {title} {{Rethinking
  $\ensuremath{\alpha}\text{\ensuremath{-}}{\mathrm{RuCl}}_{3}$}},}\ }\href
  {\doibase 10.1103/PhysRevResearch.2.033011} {\bibfield  {journal} {\bibinfo
  {journal} {Phys. Rev. Research}\ }\textbf {\bibinfo {volume} {2}},\ \bibinfo
  {pages} {033011} (\bibinfo {year} {2020})}\BibitemShut {NoStop}%
\bibitem [{\citenamefont {Banerjee}\ \emph {et~al.}(2017)\citenamefont
  {Banerjee}, \citenamefont {Yan}, \citenamefont {Knolle}, \citenamefont
  {Bridges}, \citenamefont {Stone}, \citenamefont {Lumsden}, \citenamefont
  {Mandrus}, \citenamefont {Tennant}, \citenamefont {Moessner},\ and\
  \citenamefont {Nagler}}]{Banerjee1055}%
  \BibitemOpen
  \bibfield  {author} {\bibinfo {author} {\bibfnamefont {Arnab}\ \bibnamefont
  {Banerjee}}, \bibinfo {author} {\bibfnamefont {Jiaqiang}\ \bibnamefont
  {Yan}}, \bibinfo {author} {\bibfnamefont {Johannes}\ \bibnamefont {Knolle}},
  \bibinfo {author} {\bibfnamefont {Craig~A.}\ \bibnamefont {Bridges}},
  \bibinfo {author} {\bibfnamefont {Matthew~B.}\ \bibnamefont {Stone}},
  \bibinfo {author} {\bibfnamefont {Mark~D.}\ \bibnamefont {Lumsden}}, \bibinfo
  {author} {\bibfnamefont {David~G.}\ \bibnamefont {Mandrus}}, \bibinfo
  {author} {\bibfnamefont {David~A.}\ \bibnamefont {Tennant}}, \bibinfo
  {author} {\bibfnamefont {Roderich}\ \bibnamefont {Moessner}}, \ and\ \bibinfo
  {author} {\bibfnamefont {Stephen~E.}\ \bibnamefont {Nagler}},\ }\bibfield
  {title} {\enquote {\bibinfo {title} {{Neutron scattering in the proximate
  quantum spin liquid
  $\ensuremath{\alpha}\text{\ensuremath{-}}{\mathrm{RuCl}}_{3}$}},}\ }\href
  {\doibase 10.1126/science.aah6015} {\bibfield  {journal} {\bibinfo  {journal}
  {Science}\ }\textbf {\bibinfo {volume} {356}},\ \bibinfo {pages} {1055--1059}
  (\bibinfo {year} {2017})}\BibitemShut {NoStop}%
\bibitem [{\citenamefont {Ran}\ \emph {et~al.}(2022)\citenamefont {Ran},
  \citenamefont {Wang}, \citenamefont {Bao}, \citenamefont {Cai}, \citenamefont
  {Shangguan}, \citenamefont {Ma}, \citenamefont {Wang}, \citenamefont {Dong},
  \citenamefont {膶erm谩k}, \citenamefont {Schneidewind}, \citenamefont {Meng},
  \citenamefont {Lu}, \citenamefont {Yu}, \citenamefont {Li},\ and\
  \citenamefont {Wen}}]{Kejing_Ran:27501}%
  \BibitemOpen
  \bibfield  {author} {\bibinfo {author} {\bibfnamefont {Kejing}\ \bibnamefont
  {Ran}}, \bibinfo {author} {\bibfnamefont {Jinghui}\ \bibnamefont {Wang}},
  \bibinfo {author} {\bibfnamefont {Song}\ \bibnamefont {Bao}}, \bibinfo
  {author} {\bibfnamefont {Zhengwei}\ \bibnamefont {Cai}}, \bibinfo {author}
  {\bibfnamefont {Yanyan}\ \bibnamefont {Shangguan}}, \bibinfo {author}
  {\bibfnamefont {Zhen}\ \bibnamefont {Ma}}, \bibinfo {author} {\bibfnamefont
  {Wei}\ \bibnamefont {Wang}}, \bibinfo {author} {\bibfnamefont {Zhao-Yang}\
  \bibnamefont {Dong}}, \bibinfo {author} {\bibfnamefont {P.}~\bibnamefont
  {膶erm谩k}}, \bibinfo {author} {\bibfnamefont {A.}~\bibnamefont
  {Schneidewind}}, \bibinfo {author} {\bibfnamefont {Siqin}\ \bibnamefont
  {Meng}}, \bibinfo {author} {\bibfnamefont {Zhilun}\ \bibnamefont {Lu}},
  \bibinfo {author} {\bibfnamefont {Shun-Li}\ \bibnamefont {Yu}}, \bibinfo
  {author} {\bibfnamefont {Jian-Xin}\ \bibnamefont {Li}}, \ and\ \bibinfo
  {author} {\bibfnamefont {Jinsheng}\ \bibnamefont {Wen}},\ }\bibfield  {title}
  {\enquote {\bibinfo {title} {{Evidence for Magnetic Fractional Excitations in
  a Kitaev Quantum-Spin-Liquid Candidate
  $\ensuremath{\alpha}\text{\ensuremath{-}}{\mathrm{RuCl}}_{3}$}},}\ }\href
  {\doibase 10.1088/0256-307X/39/2/027501} {\bibfield  {journal} {\bibinfo
  {journal} {Chin. Phys. Lett.}\ }\textbf {\bibinfo {volume} {39}},\ \bibinfo
  {eid} {027501} (\bibinfo {year} {2022})}\BibitemShut {NoStop}%
\bibitem [{\citenamefont {Wu}\ \emph {et~al.}(2016)\citenamefont {Wu},
  \citenamefont {Deng}, \citenamefont {Gardner}, \citenamefont {Vorderwisch},
  \citenamefont {Li}, \citenamefont {Yano}, \citenamefont {Peng},\ and\
  \citenamefont {Imamovic}}]{Wu_2016}%
  \BibitemOpen
  \bibfield  {author} {\bibinfo {author} {\bibfnamefont {C.-M.}\ \bibnamefont
  {Wu}}, \bibinfo {author} {\bibfnamefont {G.}~\bibnamefont {Deng}}, \bibinfo
  {author} {\bibfnamefont {J.S.}\ \bibnamefont {Gardner}}, \bibinfo {author}
  {\bibfnamefont {P.}~\bibnamefont {Vorderwisch}}, \bibinfo {author}
  {\bibfnamefont {W.-H.}\ \bibnamefont {Li}}, \bibinfo {author} {\bibfnamefont
  {S.}~\bibnamefont {Yano}}, \bibinfo {author} {\bibfnamefont {J.-C.}\
  \bibnamefont {Peng}}, \ and\ \bibinfo {author} {\bibfnamefont
  {E.}~\bibnamefont {Imamovic}},\ }\bibfield  {title} {\enquote {\bibinfo
  {title} {{{SIKA}{\textemdash}the multiplexing cold-neutron triple-axis
  spectrometer at {ANSTO}}},}\ }\href {\doibase 10.1088/1748-0221/11/10/p10009}
  {\bibfield  {journal} {\bibinfo  {journal} {J. Instrum.}\ }\textbf {\bibinfo
  {volume} {11}},\ \bibinfo {pages} {P10009} (\bibinfo {year}
  {2016})}\BibitemShut {NoStop}%
\bibitem [{sm()}]{sm}%
  \BibitemOpen
  \href@noop {} {}\bibinfo {note} {See Supplementary Materials at URL for
  additional inelastic neutron scattering data.}\BibitemShut {Stop}%
\bibitem [{\citenamefont {Ma}\ \emph {et~al.}(2021)\citenamefont {Ma},
  \citenamefont {Dong}, \citenamefont {Wang}, \citenamefont {Zheng},
  \citenamefont {Ran}, \citenamefont {Bao}, \citenamefont {Cai}, \citenamefont
  {Shangguan}, \citenamefont {Wang}, \citenamefont {Boehm}, \citenamefont
  {Steffens}, \citenamefont {Regnault}, \citenamefont {Wang}, \citenamefont
  {Su}, \citenamefont {Yu}, \citenamefont {Liu}, \citenamefont {Li},\ and\
  \citenamefont {Wen}}]{PhysRevB.104.224433}%
  \BibitemOpen
  \bibfield  {author} {\bibinfo {author} {\bibfnamefont {Zhen}\ \bibnamefont
  {Ma}}, \bibinfo {author} {\bibfnamefont {Zhao-Yang}\ \bibnamefont {Dong}},
  \bibinfo {author} {\bibfnamefont {Jinghui}\ \bibnamefont {Wang}}, \bibinfo
  {author} {\bibfnamefont {Shuhan}\ \bibnamefont {Zheng}}, \bibinfo {author}
  {\bibfnamefont {Kejing}\ \bibnamefont {Ran}}, \bibinfo {author}
  {\bibfnamefont {Song}\ \bibnamefont {Bao}}, \bibinfo {author} {\bibfnamefont
  {Zhengwei}\ \bibnamefont {Cai}}, \bibinfo {author} {\bibfnamefont {Yanyan}\
  \bibnamefont {Shangguan}}, \bibinfo {author} {\bibfnamefont {Wei}\
  \bibnamefont {Wang}}, \bibinfo {author} {\bibfnamefont {M.}~\bibnamefont
  {Boehm}}, \bibinfo {author} {\bibfnamefont {P.}~\bibnamefont {Steffens}},
  \bibinfo {author} {\bibfnamefont {L.-P.}\ \bibnamefont {Regnault}}, \bibinfo
  {author} {\bibfnamefont {Xiao}\ \bibnamefont {Wang}}, \bibinfo {author}
  {\bibfnamefont {Yixi}\ \bibnamefont {Su}}, \bibinfo {author} {\bibfnamefont
  {Shun-Li}\ \bibnamefont {Yu}}, \bibinfo {author} {\bibfnamefont {Jun-Ming}\
  \bibnamefont {Liu}}, \bibinfo {author} {\bibfnamefont {Jian-Xin}\
  \bibnamefont {Li}}, \ and\ \bibinfo {author} {\bibfnamefont {Jinsheng}\
  \bibnamefont {Wen}},\ }\bibfield  {title} {\enquote {\bibinfo {title}
  {{Disorder-induced broadening of the spin waves in the triangular-lattice
  quantum spin liquid candidate ${\mathrm{YbZnGaO}}_{4}$}},}\ }\href {\doibase
  10.1103/PhysRevB.104.224433} {\bibfield  {journal} {\bibinfo  {journal}
  {Phys. Rev. B}\ }\textbf {\bibinfo {volume} {104}},\ \bibinfo {pages}
  {224433} (\bibinfo {year} {2021})}\BibitemShut {NoStop}%
\bibitem [{\citenamefont {Li}\ \emph {et~al.}(2020)\citenamefont {Li},
  \citenamefont {Qu}, \citenamefont {Zhang}, \citenamefont {Jia}, \citenamefont
  {Gong}, \citenamefont {Qi},\ and\ \citenamefont
  {Li}}]{PhysRevResearch.2.043015}%
  \BibitemOpen
  \bibfield  {author} {\bibinfo {author} {\bibfnamefont {Han}\ \bibnamefont
  {Li}}, \bibinfo {author} {\bibfnamefont {Dai-Wei}\ \bibnamefont {Qu}},
  \bibinfo {author} {\bibfnamefont {Hao-Kai}\ \bibnamefont {Zhang}}, \bibinfo
  {author} {\bibfnamefont {Yi-Zhen}\ \bibnamefont {Jia}}, \bibinfo {author}
  {\bibfnamefont {Shou-Shu}\ \bibnamefont {Gong}}, \bibinfo {author}
  {\bibfnamefont {Yang}\ \bibnamefont {Qi}}, \ and\ \bibinfo {author}
  {\bibfnamefont {Wei}\ \bibnamefont {Li}},\ }\bibfield  {title} {\enquote
  {\bibinfo {title} {{Universal thermodynamics in the Kitaev fractional
  liquid}},}\ }\href {\doibase 10.1103/PhysRevResearch.2.043015} {\bibfield
  {journal} {\bibinfo  {journal} {Phys. Rev. Research}\ }\textbf {\bibinfo
  {volume} {2}},\ \bibinfo {pages} {043015} (\bibinfo {year}
  {2020})}\BibitemShut {NoStop}%
\end{thebibliography}
%

\end{document}